\documentclass[12pt]{article}

\pdfoutput=1

\usepackage{amsmath,amssymb,amsfonts}
\usepackage{color}
\definecolor{darkblue}{rgb}{0.1,0.1,.7}
\usepackage[colorlinks, linkcolor=darkblue, citecolor=darkblue, urlcolor=darkblue, linktocpage]{hyperref} 
\usepackage[square, comma, compress,numbers]{natbib}
\usepackage[]{amsmath}
\usepackage[]{graphicx}
\usepackage[]{latexsym}
\usepackage{geometry}
\usepackage{amscd}
\usepackage[all,cmtip]{xy}
\usepackage{mathrsfs}
\usepackage[margin=10pt,font=small,labelfont=bf]{caption}
\numberwithin{equation}{section}

\newcommand{\reef}[1]{(\ref{#1})}
\newcommand{\be}{\begin{equation}}
\newcommand{\ee}{\end{equation}}
\newcommand{\bea}{\begin{eqnarray}}
\newcommand{\eea}{\end{eqnarray}}

\newcommand{\SR}[1]{{ #1}}

\newcommand{\eps}{\epsilon}
\def\beq{\begin{equation}} 
\def\eeq{\end{equation}} 
\def\spin{\sigma}
\def\en{\varepsilon} 
\def\del {\partial} 
\def\nn{\nonumber} 
\def\bZ {\mathbb{Z}} 
\def\calO {{\cal O}} 
 
\def\calF {{\cal F}} 
\def\spin{\sigma}
\def\en{\varepsilon} 
\def\bZ {\mathbb{Z}} 
\def\half{\textstyle\frac 12}

\begin{document}

\vspace*{-.6in} \thispagestyle{empty}
\begin{flushright}
LPTENS--12/07
\end{flushright}
\vspace{.2in} {\Large
\begin{center}
{\bf Solving the 3D Ising Model \\\vspace{.1in} with the Conformal Bootstrap}
\end{center}
}
\vspace{.2in}
\begin{center}
{\bf Sheer El-Showk$^{a}$, 
Miguel F. Paulos$^{b}$, 
David Poland$^{c}$,}\\\vspace{.2in} 
{\bf Slava Rychkov$^{d}$, 
David Simmons-Duffin$^{e}$, 
Alessandro Vichi$^{f}$}
\\
\vspace{.2in} 
$^a$ {\it Institut de Physique Th\'eorique CEA Saclay, CNRS-URA 2306, \\ 91191 Gif sur Yvette, France}
\\\vspace{.2in} 
$^b$ {\it Laboratoire de Physique Th\'{e}orique et Hautes Energies, CNRS UMR 7589,\\ Universit\'e Pierre et Marie Curie, 4 place Jussieu, 75252 Paris Cedex 05, France}
\\\vspace{.2in} 
$^c$ {\it School of Natural Sciences, Institute for Advanced Study, \\Princeton, New Jersey 08540, USA}
\\\vspace{.2in} 
$^d$ {\it  Facult\'{e} de Physique, Universit\'{e} Pierre et Marie Curie\\
\& Laboratoire de Physique Th\'{e}orique, \'{E}cole Normale Sup\'{e}rieure, Paris, France}
\\\vspace{.2in} 
$^e$ {\it Jefferson Physical Laboratory, Harvard University, Cambridge, MA 02138, USA}
\\\vspace{.2in} 
$^f$ {\it Theoretical Physics Group, Ernest Orlando Lawrence Berkeley National Laboratory\\
 \& Center for Theoretical Physics, University of California, Berkeley, CA 94720, USA}
\end{center}

\vspace{.2in}

\begin{abstract}
We study the constraints of crossing symmetry and unitarity in general 3D Conformal Field Theories.  In doing so we derive new results for conformal blocks appearing in four-point functions of scalars and present an efficient method for their computation in arbitrary space-time dimension.  Comparing the resulting bounds on operator dimensions and OPE coefficients in 3D to known results, we find that the 3D Ising model lies at a corner point on the boundary of the allowed parameter space. We also derive general upper bounds on the dimensions of higher spin operators, relevant in the context of theories with weakly broken higher spin symmetries.

\end{abstract}
\vspace{.2in}
\vspace{.3in}
\hspace{0.7cm} March 2012

\newpage
\setcounter{page}{1}

\tableofcontents

\newpage


\section{Introduction}

This paper is the first in a series of works which will conceivably lead to a solution of the Conformal Field Theory (CFT) describing the three dimensional (3D) Ising model at the critical temperature. Second-order phase transitions in a number of real-world systems are known to belong to the same universality class: most notably liquid-vapor transitions and transitions in binary fluids and uniaxial magnets.

Field-theoretical descriptions of critical phenomena and computations of critical exponents have a long tradition \cite{Pelissetto:2000ek}. One well-known approach to this problem is the $\eps$-expansion \cite{Wilson:1971dc}. In this method the critical exponents are computed using the usual, perturbative field theory in $D=4-\eps$ dimensions, and the physically interesting case of $D=3$ is obtained by extrapolating to $\eps=1$. The obtained series in $\eps$ are divergent and need to be resummed. Apart from small ambiguities, the final results for the critical exponents agree well with experiments and with a host of other approximation techniques (high-temperature expansion, Monte-Carlo simulations, etc.).

In this paper we will develop an alternative method for determining critical exponents in $D=3$, based on Polyakov's hypothesis of conformal invariance of critical fluctuations \cite{Polyakov:1970xd}, which was a major motivation for the development of Conformal Field Theory.  CFT methods have been extremely fruitful in $D=2$, allowing one to solve many models of critical behavior \cite{Belavin:1984vu}. The novelty of our project is to apply them in $D=3$. The existing quantitative approaches to critical phenomena in $D=3$ do not take full advantage of conformal invariance.

The CFT describing the 3D Ising model at criticality is not known to possess any additional symmetry apart from conformal invariance and $\bZ_2$ invariance. For this reason we will be able to rely only on the most general properties of conformal theories.
The study of such general properties goes back to the 1970s. The required fundamental concepts are the classification of primary operators, the conformally-invariant operator product expansion, conformal blocks, and the idea of the nonperturbative conformal bootstrap, which were introduced in the work of Mack and Salam \cite{Mack:1969rr}, Ferrara, Gatto, Grillo and Parisi \cite{Ferrara:1971vh,Ferrara:1973vz,Ferrara:1973yt,Ferrara:1974nf,Ferrara:1974ny,Ferrara:1974pt} and Polyakov \cite{Polyakov:1974gs}. In addition, we will need explicit expressions for the conformal blocks. Here we will be able to rely on the recent work of Dolan and Osborn \cite{DO1,DO2,DO3}.

While most of these ingredients were understood many years ago, until recently it was not known how to put them together in order to extract dynamical information about CFTs. This important know-how was developed in a series of recent papers \cite{Rattazzi:2008pe,Rychkov:2009ij,Caracciolo:2009bx,Poland:2010wg,Rattazzi:2010gj,Rattazzi:2010yc,Vichi:2011ux,Poland:2011ey}. That work was largely motivated by particle physics (in particular the theory of electroweak symmetry breaking) and concerned CFT in $D=4$. However, the time is now ripe to transfer these techniques to $D=3$. The cases $D=3$ and $D=4$ are similar in that the conformal algebra has finitely many generators (unlike in $D=2$ where it has an infinite-dimensional extension, the Virasoro algebra).

This paper is structured as follows.  In Section \ref{sec:operatorcontent} we review what is known about the operator content of the 3D Ising model.  In Section \ref{sec:bootstrap} we discuss the conformal bootstrap approach to studying 3D CFTs, and in Section \ref{sec:blocks} we present an efficient method for computing the conformal partial waves appearing in four-point functions of scalars for CFTs in any dimension (including $D=3$).  In Section \ref{sec:bounds} we present bounds on 3D CFTs that follow from crossing symmetry and compare them to what is known about the 3D Ising model.  Finally, we discuss our results and future directions for this program in Section \ref{sec:discussion}.

\section{Operator Content of the 3D Ising Model}
\label{sec:operatorcontent}

We assume that the reader is familiar with the basic facts about the Ising model and the critical phenomena in general, see \cite{Patashinsky:1979yx,Cardy:1996xt,ZinnJustin:2002ru,Kleinert,Pelissetto:2000ek}. 

In this paper, we will be aiming for a solution of the 3D Ising model \emph{in the continuum limit} and \emph{at the critical temperature} $T=T_c$. While the 2D Ising model was solved exactly \emph{on the lattice} and \emph{for any temperature} by Onsager and Kaufman in the 1940's, the 3D lattice case has resisted all attempts for an exact solution. Istrail 
\cite{Istrail} proved in 2000 that solving the 3D Ising model on the lattice is an NP-complete problem. However, this theorem does not exclude the possibility of finding a solution in the continuum limit.

The standard way to think about the continuum theory is in terms of local operators (or fields). At $T=T_c$, the theory has scale (and, as we discuss below, conformal) invariance, and each operator is characterized by its scaling dimension $\Delta$ and $O(3)$ spin. The operators of spin higher than 1 are traceless symmetric tensors. 

In Table \ref{tab:dims} we list a few notable local operators, which split into odd and even sectors under the global $\bZ_2$ symmetry (the Ising spin flip). The operators $\spin$ and $\en$ are the lowest dimension $\bZ_2$-odd and even scalars respectively---these are the continuum space versions of the Ising spin and of the product of two neighboring spins on the lattice. The two next-to-lowest scalars in each $\bZ_2$-sector are called $\spin'$ and $\en'$.  Their dimensions are related to the irrelevant critical exponents $\omega_A$ and $\omega$ measuring corrections to scaling. The operator $\en''$ is analogously related to the next-to-leading $\bZ_2$-even irrelevant exponent $\omega_2$. The stress tensor $T_{\mu\nu}$ has spin 2 and, as a consequence of being conserved, canonical dimension $\Delta_T=3$. The lowest-dimension spin 4 operator $C_{\mu\nu\kappa\lambda}$ has a small anomalous dimension, related to the critical exponent $\omega_{\rm NR}$ measuring effects of rotational symmetry breaking on the cubic lattice.
  
\begin{table}[htdp]
\begin{center}
\begin{tabular}{|c|c|c|l|l|}
\hline
Operator & Spin $l$ & $\bZ_2$ & $\Delta$ & Exponent \\
\hline
\hline
$\spin$ & 0 & $-$ & 0.5182(3) & $\Delta=1/2+\eta/2$\\
$\spin'$ & 0 & $-$ & $\gtrsim 4.5$ & $\Delta=3+ \omega_A$\\
$\en$ & 0 & $+$ & 1.413(1) & $\Delta=3-1/\nu$\\
$\en'$ & 0 & $+$ & 3.84(4) & $\Delta=3+\omega$\\
$\en''$ & 0 & $+$ & 4.67(11) & $\Delta=3+\omega_2$\\
$T_{\mu\nu}$ & 2 & $+$ & 3 & n/a\\
$C_{\mu\nu\kappa\lambda}$ & 4 & $+$ & 5.0208(12) & $\Delta= 3+\omega_{\rm NR}$\\
\hline
\end{tabular}
\end{center}
\caption{Notable low-lying operators of the 3D Ising model at criticality.}
\label{tab:dims}
\end{table}%

The approximate values of operator dimensions given in the table have been determined from a variety of theoretical techniques, most notably the $\eps$-expansion, high temperature expansion, and Monte-Carlo simulations; see p.~47 of Ref.~\cite{Pelissetto:2000ek} for a summary. The achieved precision is rather impressive for the lowest operator in each class, but quickly gets worse for the higher fields.  While ultimately we would like to beat the old methods, it would be unwise to completely dismiss this known information and restart from scratch. Rather, we will be using it for guidance while sharpening our own methods. 

Among the old techniques, the $\eps$-expansion of Wilson and Fisher \cite{Wilson:1971dc} deserves a separate comment. The well-known idea of this approach is that the 3D Ising critical point and the 4D free scalar theory can be connected by a line of fixed points by allowing the dimension of space to vary continuously between 3 and 4. For $D=4-\eps$, the Wilson-Fisher fixed point is weakly coupled and the dimensions of local operators can be expanded order-by-order in $\eps$. For the most important operators, like $\spin$ and $\en$, these expansions have been extended to terms of order as high as $\eps^5$ \cite{ZinnJustin:2002ru}, requiring a five-loop perturbative field theory computation. However, as often happens in perturbation theory, the resulting series are only asymptotic. For the physically interesting case $\eps=1$, their divergent nature already starts to show after the first couple of terms. Nevertheless, after appropriate resummation the $\eps$-expansion produces results in agreement with the other methods. So its basic hypothesis must be right, and can give useful qualitative information about the 3D Ising operator spectrum, even where accurate quantitative computations are missing.

It is now time to bring up the conformal invariance of the critical point, conjectured by Polyakov \cite{Polyakov:1970xd}. This symmetry is left unused in the RG calculations leading to the $\eps$-expansion, and in most other existing techniques.\footnote{Conformal invariance has been used in studies of critical $O(N)$ models in the large $N$ limit \cite{Lang:1991kp,Petkou:1994ad}.} 
This is because it only emerges at the critical point; it's not present along the flow. Conformal invariance seems to be a generic feature of criticality, but why exactly is not fully understood \cite{Polchinski:1987dy}. Recently there has been a renewed interest in the question of whether there exist interesting scale invariant but not conformal systems \cite{Dorigoni:2009ra, ElShowk:2011gz, Antoniadis:2011gn, Fortin:2011sz,Nakayama:2011tk,Fortin:2012ic}. We will simply assume as a working hypothesis that the 3D Ising critical point is conformal. 

A nice experimental test of conformal invariance would be to measure the three-point function $\langle \spin(x) \spin(y) \en(z)\rangle$ on the lattice, to see if its functional form agrees with the one fixed by conformal symmetry \cite{Polyakov:1970xd}. We do not know if this has been done.

Using 3D conformal invariance, local operators can be classified into primaries and descendants \cite{Mack:1969rr}. 
The primaries\footnote{These are usually called quasi-primaries in 2D CFTs.} transform homogeneously under the finite-dimensional conformal group, while the descendants are derivatives of primaries and transform accordingly. All operators listed in Table~\ref{tab:dims} are primaries. This is obvious for $\spin$ and $\en$---the lowest dimension scalars in each $\bZ_2$-symmetry class. That $\spin'$, $\en'$, $\en'', C_{\mu\nu\kappa\lambda}$ are all primaries and not derivative operators follows from the fact that they are associated with corrections to scaling, while adding a derivative operator to the Lagrangian has no effect. Finally, the stress tensor is always a primary. 

It can be seen that all operators in Table~\ref{tab:dims} have non-negative anomalous dimensions (by which we mean the difference between the operator dimension and the dimension of the lowest 3D free scalar theory operator with the same quantum numbers). This is not accidental, but is related to reflection positivity, which is the Euclidean space version of unitarity. Primaries in reflection positive (or unitary) CFTs are known to have non-negative anomalous dimensions \cite{Ferrara:1974pt,Mack:1975je,Metsaev:1995re,Minwalla:1997ka,Grinstein:2008qk}:
\begin{gather}
\Delta\ge D/2-1 \quad (l=0)\,,\qquad
\Delta\ge l+D-2 \quad (l\ge1)\,.
\label{eq:ubounds}
\end{gather}
The 3D Ising model is reflection positive on the lattice \cite{Glimm:1981xz}, and this property is inherited in the continuum limit, so that the `unitarity bounds' \reef{eq:ubounds} are respected. 

Can conformal symmetry be used to \emph{determine} the local operator dimensions rather than to interpret the results obtained via other techniques? In 2D this was done long ago \cite{Belavin:1984vu} using the Virasoro algebra. This also justified \emph{post factum} the assumption of conformal invariance, since the critical exponents and other quantities agreed with the exact lattice solution. The Virasoro algebra does not extend to 3D, but in the next section we will describe a method which is applicable for any $D$.

\section{Conformal Bootstrap}
\label{sec:bootstrap}

Primary operators in a CFT form an algebra under the Operator Product Expansion (OPE). This means that the product of two primary operators at nearby points can be replaced inside a correlation function by a series in other local operators times coordinate-dependent coefficient functions. Schematically, the OPE of two primaries has the form:
\beq
\phi_i(x_1) \phi_j(x_2)=\sum_k f_{ijk}\, C(x_1-x_2,\del_2) \phi_k(x_2) .
\eeq
The differential operators $C$ are fixed by conformal invariance, and only primary operators need to be included in the sum on the RHS.  Here we are suppressing indices for clarity.  In general, scalar operators as well as operators of nonzero spin will appear on the RHS.  Fairly explicit expressions for the $C$'s have been known since the 70's, at least in the case when the $\phi_{i,j}$ are scalars and $\phi_k$ is a traceless symmetric tensor of arbitrary rank \cite{Ferrara:1971vh,DO1}, but we will not need them here. 

The numerical coefficients $f_{ijk}$ are called structure constants, or OPE coefficients. These numbers, along with the dimensions and spins of all primary fields, comprise the `CFT data' characterizing the algebra of local operators.

\begin{figure}[htbp]
\begin{center}
\includegraphics[scale=0.5]{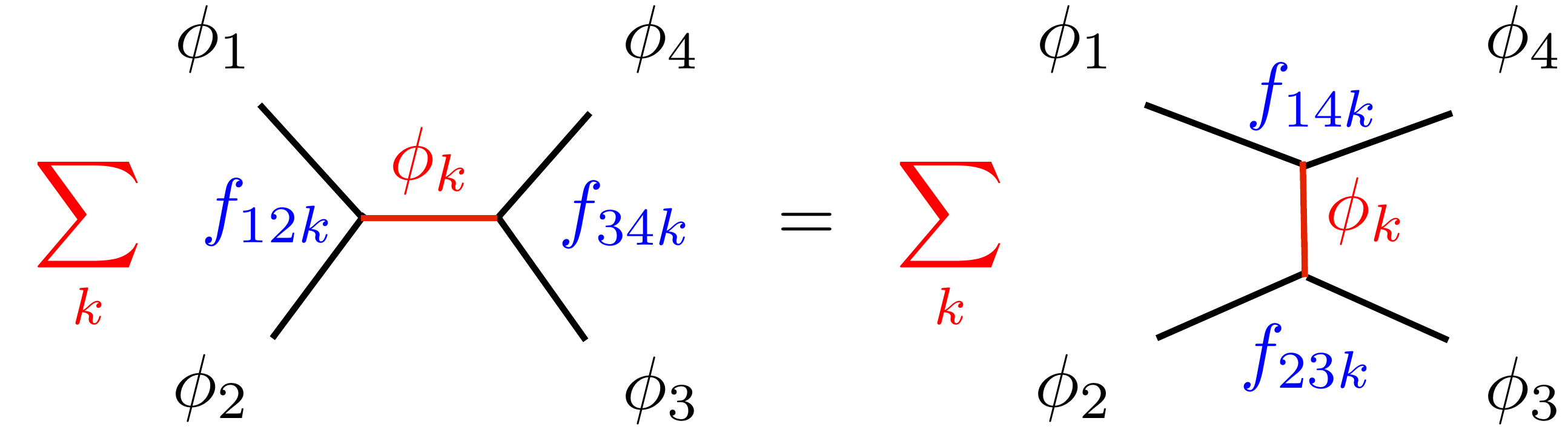}
\caption{The conformal bootstrap condition $=$ associativity of the operator algebra.}
\label{fig:ass}
\end{center}
\end{figure}

The conformal bootstrap condition \cite{Ferrara:1973yt,Polyakov:1974gs,Belavin:1984vu}, shown schematically in Fig.~\ref{fig:ass}, says that the operator algebra must be associative. In that figure we consider the correlator of four primaries 
\beq
\langle\phi_1(x_1)\phi_2(x_2)\phi_3(x_3)\phi_4(x_4)\rangle
\eeq
and use the OPE in the (12)(34) or (14)(23)-channel to reduce it to a sum of two-point functions. The answer should be the same, which gives a quadratic condition on the structure constants of the schematic form
\beq
\sum_k f_{12k}f_{34k}(\ldots)=\sum_k f_{14k}f_{23k}(\ldots)\,.
\label{eq:boot}
\eeq
The $(\ldots)$ factors are functions of coordinates $x_i$, called conformal partial waves. They are produced by acting on the two-point function of the exchanged primary field $\phi_k$ with the differential operators $C$ appearing in the OPE of two external primaries. Thus, they are also fixed by conformal invariance in terms of the dimensions and spins of the involved fields.

The dream of the conformal bootstrap is that the condition \reef{eq:boot}, when imposed on four-point functions of sufficiently many (all?) primary fields, should allow one to determine the CFT data and thus solve the CFT. Of course, there are presumably many different CFTs, and so one can expect some (discrete?) set of solutions. One of the criteria which will help us to select the solution representing the 3D Ising model is the global symmetry group, which must be $\bZ_2$.

Our method of dealing with the conformal bootstrap will require explicit knowledge of the conformal partial waves. In the next section we will gather the needed results.

\section{Conformal Blocks}
\label{sec:blocks}

In this paper we will be imposing the bootstrap condition only on four-point functions of scalars. Conformal partial waves for such correlators were introduced in \cite{Ferrara:1973vz} and further studied in \cite{Ferrara:1974nf,Ferrara:1974ny}; they were also discussed in \cite{Polyakov:1974gs}. Recently, new deep results about them were obtained in \cite{DO1,DO2,DO3}.  Significant progress in understanding non-scalar conformal partial waves was made recently in~\cite{Costa:2011dw} (building on~\cite{Costa:2011mg}), which also contains a concise introduction to the concept.  Below we'll normalize the scalar conformal partial waves as in \cite{DO3}; see Appendix \ref{app:recursion} for further details on our conventions.

Consider a correlation function of four scalar primaries $\phi_i$ of dimension $\Delta_i$, which is fixed by conformal invariance to have the form \cite{Polyakov:1970xd}
\begin{equation}
\langle\phi_{1}(x_{1})\phi_{2}(x_{2})\phi_{3}(x_{3})\phi_{4}(x_{4})\rangle
=\left(  \frac{x^2_{24}}{x^2_{14}}\right)  ^{\frac12\Delta_{12}%
}\left(  \frac{x^2_{14}}{x^2_{13}}\right)  ^{\frac12\Delta_{34}}
\frac{g(u,v)}{(x^2_{12})^{\frac12(\Delta_{1}+\Delta_{2})}(x_{34}^2)^{\frac12(\Delta_{3}+\Delta_{4})%
}}\,, \label{eq:4pt}%
\end{equation}
where $x_{ij}\equiv x_i-x_j$, $\Delta_{ij}\equiv \Delta_i-\Delta_j$, and $g(u,v)$ is a function of the conformally invariant cross-ratios
\begin{equation}
u=\frac{x_{12}^{2}x_{34}^{2}}{x_{13}^{2}x_{24}^{2}},\quad v=\frac{x_{14}%
^{2}x_{23}^{2}}{x_{13}^{2}x_{24}^{2}}.
\end{equation}
The conformal partial wave expansion in the (12)(34) channel gives a series representation for this function:
\beq
g(u,v)= \sum_\calO f_{12\calO}f_{34\calO}\, {G_{\Delta,l}(u,v)}\,,
\label{eq:CBexp}
\eeq
where the sum is over the exchanged primaries $\calO$ of dimension $\Delta$ and spin $l$ and the functions $G_{\Delta,l}(u,v)$ are called conformal blocks. We must learn to compute them efficiently.

\begin{figure}[htbp]
\begin{center}
\includegraphics[scale=0.5]{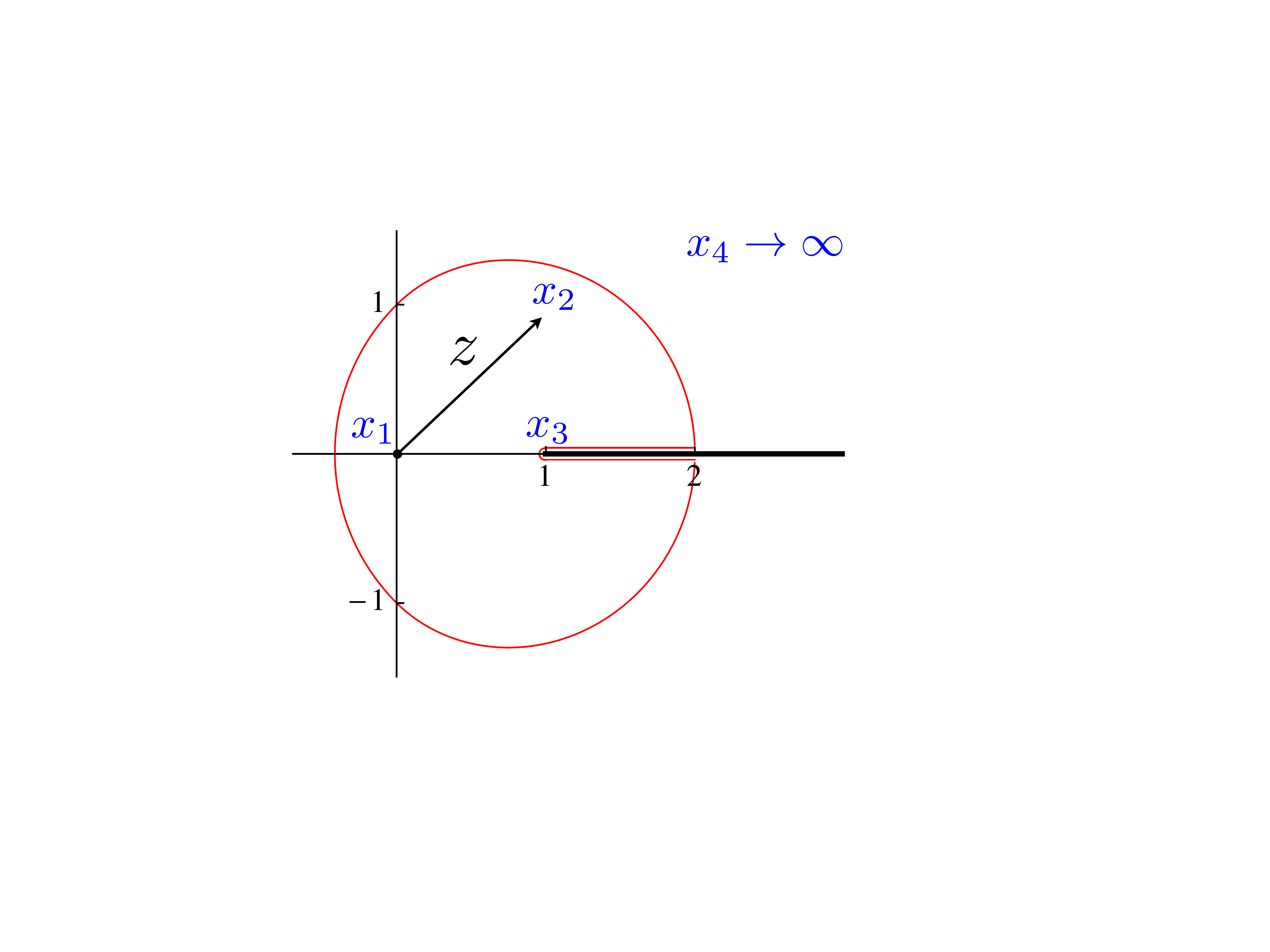}
\caption{Using conformal freedom, three operators can be fixed at $x_1=0$, $x_3=(1,0,\ldots,0)$, $x_4\to \infty$, while the fourth point  $x_2$ can be assumed to lie in the (12) plane. The variable $z$ is then the complex coordinate of $x_2$ in this plane, while $\bar z$ is its complex conjugate. Also shown: the conformal block analyticity cut (thick black), and the boundary of the absolute convergence region of the power series representation \reef{eq:doublesum} 
(thin red).}
\label{fig:z}
\end{center}
\end{figure}

In even dimensions, conformal blocks have relatively simple closed-form expressions in terms of hypergeometric functions \cite{Ferrara:1974ny,DO1,DO2,DO3}. For example, the 2D and 4D blocks are given by: 
\begin{gather}
G^{D=2}_{\Delta,l}(u,v)   =\frac12\left[ k_{\Delta+l}(z)k_{\Delta-l}(\bar{z})+(z\leftrightarrow\bar{z})\right]\,,\nn\\
G^{D=4}_{\Delta,l}(u,v)   =\frac{1}{l+1}
\frac{z\bar{z}}{z-\bar{z}}\left[ k_{\Delta+l}(z)k_{\Delta-l-2}
(\bar{z})-(z\leftrightarrow\bar{z})\right],\label{eq:DO}
\end{gather}
where
\beq
k_{\beta}(x)  \equiv x^{\beta/2}{}_{2}F_{1}\left(  \half(\beta-\Delta_{12}),\half(\beta+\Delta_{34});\beta;x\right)\,, 
\eeq
and the complex variable $z$ and its complex conjugate $\bar{z}$ are related to $u,v$ via%
\begin{equation}
u=z\bar{z},\quad v=(1-z)(1-\bar{z})\,.
 \label{eq:uvzzbar}%
\end{equation}
The meaning of the variable $z$ is explained in Fig.~\ref{fig:z}. From the known analyticity properties of ${}_{2}F_{1}$, it follows that the conformal blocks are smooth single-valued functions in the $z$ plane minus the origin and the $(1,+\infty)$ cut along the real axis. This is not accidental and should be valid for any $D$. By standard radial quantization reasoning (see \cite{Polchinski:1998rq}, Sec.~2.9), the OPE by which the conformal blocks are defined is expected to converge as long as there is a sphere separating $x_1$ and $x_2$ from $x_3$ and $x_4$. This sphere degenerates into a plane and disappears precisely when $z$ crosses the cut.

We now pass to the results for general $D$, including the case $D=3$ we are interested in, which are rather more complicated. From now on we consider only conformal blocks of four \emph{identical} scalars, so that $\Delta_{12}=\Delta_{34}=0$. In any dimension, such blocks depend only on the dimension and spin of the exchanged primary. For scalar exchange ($l=0$), conformal blocks have a double power series representation (\cite{DO1}, Eq.~(2.32)):
\beq
\label{eq:doublesum}
G_{\Delta,0}(u,v)=u^{\Delta/2}\sum_{m,n=0}^\infty
\frac{[(\Delta/2)_m (\Delta/2)_{m+n}]^2 }{m!\, n!(\Delta+1-\frac{D}{2})_m  (\Delta)_{2m+n}}
u^m(1-v)^n,
\eeq
where $(x)_n$ is the Pochhammer symbol. In this paper we will only use this representation at $z=\bar z$, in order to derive the closed form expression \reef{eq:3F2-0} given below. In principle the series converges absolutely in the region 
\beq
|1-v|<\begin{cases} 1,& 0\le u<1\,, \\
2\sqrt{u}-u, & u\ge 1\,,
\end{cases}
\eeq
whose boundary is traced in red in Fig. \ref{fig:z}. 

For exchanged operators of nonzero spin, the conformal blocks can be computed via various recursion relations. Some recursion relations previously appeared in \cite{DO1}, Eqs.~(2.30), but these will not be useful for us since they express the blocks with equal external dimensions in terms of blocks where the external dimensions differ by an integer. 

In Appendix~\ref{app:recursion}, we exhibit a recursion relation which follows from the results of Ref.~\cite{DO3} and does not require shifts in the external dimensions.  
In general, this recursion involves taking derivatives of $G_{\Delta,l}(z,\bar{z})$, which is not very easy to perform numerically.  However, along the line $z=\bar z$ the terms involving derivatives drop out and the recursion relation for the conformal block $G_{\Delta,l}(z) \equiv G_{\Delta,l}(z,z)$ becomes extremely simple:	
	\begin{multline}
	(l+D-3)(2\Delta+2-D) G_{\Delta,l}(z) \\
	=(D-2) (\Delta+l-1) G_{\Delta,l-2}(z)
	+\frac{2-z}{2z} \,(2l+D-4)(\Delta-D+2)
	G_{\Delta+1,l-1}(z)\\
	-\frac{\Delta(2l+D-4)(\Delta+2-D)(\Delta+3-D)(\Delta-l-D+4)^2}
	{16(\Delta+1-\frac{D}{2})(\Delta-\frac{D}{2}+2)(l-\Delta+D-5)(l-\Delta+D-3)} G_{\Delta+2,l-2}(z) . 
	\label{eq:recz=zb}
	\end{multline}
	
This recursion relation can easily compute all conformal blocks along the $z=\bar z$ line in terms of spin 0 and 1 blocks.\footnote{\label{note:-1}This works for general $D$. In $D=3$, one can instead recurse from $G_{\Delta,0}$ and $G_{\Delta,-1}\equiv G_{\Delta,0}$, where the latter equality follows from \reef{eq:mirror}.} On the other hand, as shown in Appendix \ref{app:scalarblocks}, the spin 0 and 1 blocks along the $z=\bar z$ line can be simply expressed in terms of generalized hypergeometric functions ($\alpha\equiv D/2-1$): 
\begin{gather}
G_{\Delta,0}(z) =
\left(\frac{z^{2}}{1-z}\right)^{\Delta/2}
\, _3F_2\left({\textstyle\frac{\Delta}2,\frac{\Delta}2,\frac{\Delta
   }{2}-\alpha ;\frac{\Delta+1 }{2},\Delta -\alpha}
   ;\frac{z^2}{4 (z-1)}\right)\,,  
\label{eq:3F2-0}  \\
G_{\Delta,1}(z) =\frac{2-z }{2 z} \left(\frac{z^2}{1-z}\right)^{\frac{\Delta +1}{2}}
\, _3F_2\left({\textstyle\frac{\Delta+1 }{2},\frac{\Delta+1
   }{2},\frac{\Delta+1 }{2}-\alpha ;\frac{\Delta
   }{2}+1,\Delta -\alpha };\frac{z^2}{4 (z-1)}\right)\,.
   \label{eq:3F2-1} 
\end{gather}

These explicit expressions, together with the recursion relation \reef{eq:recz=zb}, solve the problem of finding conformal blocks along the $z=\bar z$ line. What about $z\ne \bar z$?
We should explain that in our numerical implementation of conformal bootstrap we will not actually use the values of conformal blocks at generic values of $z$. Instead, we will Taylor-expand the conformal bootstrap condition around the point $z=\bar z=1/2$. This is an approach which proved efficient in prior work in 4D and 2D, and we will pursue it here as well. So, we will have to evaluate derivatives of conformal blocks at the point $z=\bar z=1/2$, both along and transverse to the $z=\bar z$ line.

Now, derivatives along the $z=\bar z$ line will be evaluated as follows. For the spin 0 and 1 conformal blocks we can take advantage of the fact that the $_3 F_2$ hypergeometric functions satisfy a third-order differential equation:
	\bea
	\left(x \hat D_{a_1} \hat D_{a_2} \hat D_{a_3}-\hat D_{0}\hat D_{b_1-1}\hat D_{b_2-1}\right)\, _3 F_2(a_1,a_2,a_3;b_1,b_2;x)=0\,,
	\eea
where $\hat D_c\equiv x \partial_x+c$. This equation can be used to obtain recursion relations which express the third-order and higher derivatives of the spin 0 and 1 blocks in terms of their first and second derivatives. The values of the latter derivatives at $z=\bar z=1/2$ will be tabulated as a function of $\Delta$.

Derivatives of the higher spin blocks are then computed using the recursion relations following from \reef{eq:recz=zb}. This completely settles the question of obtaining derivatives along the $z=\bar z$ line. 

In order to obtain the derivatives transverse to the $z=\bar z$ line, we'll take advantage of the fact that conformal partial waves are eigenfunctions of the quadratic Casimir operator of the conformal group, which implies that conformal blocks satisfy a second-order differential equation \cite{DO2}:
	\bea
	\mathcal D G_{\Delta,l}(z,\bar z)=\half C_{\Delta,l} G_{\Delta,l}(z,\bar z)\,,
	\label{eq:Cas}
	\eea
where $C_{\Delta,l}\equiv \Delta(\Delta-D)+l (l+D-2)$ and
	\bea
	\mathcal D \equiv (1-z)z^2 \partial_z^2-
	\left[z^2-(D-2) \frac{z \bar z (1-z)}{z-\bar z}\right] \partial_z+ (z \leftrightarrow \bar z)\,.
	\eea
Let us now make a change of variables:
\beq 
\label{eq:abdefinition}
z=(a+\sqrt{b})/2,\quad \bar z=(a-\sqrt{b})/2.
\eeq 
The point $z=\bar z=1/2$ which interests us corresponds to $a=1$, $b=0$. Moreover, since conformal blocks are symmetric in $z\leftrightarrow \bar z$, their power series expansion away from the $z=\bar z$ line will contain only even powers of $(z-\bar z)$, and hence integer powers of $b$. In the new variables the differential operator $\mathcal D$ takes the form
\begin{align}
\mathcal 
D =\, & (2-a) a^2\left[\frac{1}{2} (D-1) \del_b + b\, \del_b^2\right] \nn\\
 &\,+\left\{(2-3 a) b^2 \del_b^2 + \left[\frac{1}{2} (D-9)a-a (3 a-4)   \del_a- D+3\right] b\, \del_b -\frac{1}{4} Da^2 \del_a +\frac{1}{4} (2-a) a^2 \del_a^2\right\} \nn\\
  &\,+\left\{-b^2 \del_a \del_b + \left[\frac{1}{4} (D-4) \del_a+\frac{1}{4} (2-3 a) \del_a^2\right]b\right\}\,,
   \label{eq:Casab}
   \end{align}
where the terms have been grouped into lines according to how they change the power of $b$. The first line contains the \emph{leading terms}, which lower the power of $b$ by one unit. Notice that the leading terms generate a nonvanishing coefficient when acting on any positive power of $b$, as long as $0<a<2$ (which corresponds to $0<z<1$). 
Thus, in a neighborhood of this interval the Casimir differential equation \reef{eq:Cas} can be solved \emph{\`a la} Cauchy-Kovalevskaya, recursively in a power series expansion in $b$ using the known conformal blocks at $b=0$ as a boundary value. 

Let us denote the $\del_a^m\del_b^n$ derivative of the conformal block evaluated at $z=\bar z=1/2$ by $h_{m,n}$. Since we know the conformal blocks along the $z=\bar z$ line, we can compute all the derivatives $h_{m,0}$. On the other hand, the Cauchy-Kovalevskaya argument above implies that there will be a recursion relation for $h_{m,n}$ (with $n>0$) in terms of $h_{m,n}$ with lower values of $n$. The recursion relation is given in Appendix \ref{app:CKrec} and has the general structure: 
\begin{align}
h_{m,n}=\sum_{m'\le m-1} m (\ldots) h_{m',n} +\sum_{m'\le m+2}\left[ (\ldots) h_{m',n-1} + (n-1) (\ldots) h_{m',n-2} \right]\,.
\end{align}
The appearance of $m'$ up to $m+2$ is related to the fact that, the Casimir equation being second-order, derivatives of up to second-order in $a$ appear in the RHS of \reef{eq:Casab}.  The first term being proportional to $m$ ensures that $h_{0,n}$ terms generated by repeatedly applying the recursion are eventually reduced to $h_{1,0}$ and $h_{0,0}$.  This recursion then solves the problem of computing the conformal block derivatives transverse to the line $z=\bar z$.
	
\section{Bounds and Consequences for the 3D Ising Model}
\label{sec:bounds}

In this section we will use the bootstrap equations discussed above in order to derive rigorous bounds on 3D CFTs.  When comparing these bounds to the 3D Ising model, we'll focus on constraints coming from the four-point function of the Ising spin operator $\langle \spin\spin\spin\spin\rangle$.  The conformal block expansion of this four-point function has the form
\beq
g(u,v)=\sum p_{\Delta,l} G_{\Delta,l}(u,v)\,,
\quad p_{\Delta,l}\equiv f_{\Delta,l}^2\ge 0\,,
\eeq
where the sum runs over the dimensions and spins of all primary operators appearing in the $\sigma\times\sigma$ OPE. This OPE contains all of the $\bZ_2$-even operators listed in Table \ref{tab:dims}, in addition to infinitely many other even-spin operators. Note that odd-spin operators cannot appear because of Bose symmetry. The coefficients $p_{\Delta,l}$ appearing in the conformal block expansion are squares of the OPE coefficients, and are thus constrained to be positive.

The conformal bootstrap equation \reef{eq:boot} takes a particularly simple form for this correlator, since the (12)(34) and (14)(23) channel involve the same OPE coefficients. It can be stated as a crossing symmetry constraint on the function $g(u,v)$:
\beq
v^{\Delta_\spin} g(u,v)=u^{\Delta_\spin}  g(v,u)\,.
\eeq
Substituting the conformal block decomposition, we get an equation
\beq
u^{\Delta_\spin}-v^{\Delta_\spin}=\sum\nolimits' p_{\Delta,l}\left[ v^{\Delta_\spin}  G_{\Delta,l}(u,v)-u^{\Delta_\spin}  G_{\Delta,l}(v,u) \right]\,,
\label{eq:cross}
\eeq
where $\sum\nolimits'$ is the sum over all operators except the unit operator, whose contribution has been separated in the LHS.
It was shown in \cite{Rattazzi:2008pe} and confirmed in subsequent work \cite{Rychkov:2009ij,Caracciolo:2009bx,Poland:2010wg,Rattazzi:2010gj,Rattazzi:2010yc,Vichi:2011ux,Poland:2011ey} that this type of equation can be used to extract dynamical information about 4D and 2D CFTs. We will now apply the same methods in 3D.

First, we will Taylor-expand \reef{eq:cross} around the point $z=\bar z=1/2$ up to some large fixed order. That this is a reasonable point to expand around follows from the fact that it is democratic with respect to the direct and crossed channels in the conformal block decomposition: by making a conformal transformation the four points can be put at the vertices of a square.

The Taylor-expanded \reef{eq:cross} can be viewed as a finite system of linear equations (one for each Taylor coefficient) for a large (strictly speaking infinite) number of variables $p_{\Delta,l}$. A priori, there is one variable $p_{\Delta,l}$ for each pair $(\Delta,l)$ consistent with the unitarity bounds \reef{eq:ubounds}. However, one may wish to posit additional constraints on the spectrum (such as assumptions about gaps). Below we will study which of these constraints are  consistent with the existence of a solution. 

This system of linear equations should also be augmented by inequalities expressing the fact that variables $p_{\Delta,l}$ are non-negative. Fortunately, problems involving linear inequalities are almost as tractable as pure systems of linear equalities. These problems form a chapter of linear algebra called linear programming, and there exist efficient algorithms for solving them (such as Dantzig's simplex method or interior point methods). Once the additional constraints on the spectrum are specified, one can use linear programming methods to find out if the system has a solution. If the answer is negative, a CFT with such a spectrum cannot exist.  The details of our implementation of this problem are given in Appendix \ref{app:details}.

\subsection{Bounds on $\Delta_\en$}

We are now ready to start asking concrete questions about the 3D Ising CFT to which we can give unambiguous answers. The first question is as follows. Let's be agnostic about the dimension of the spin field, allowing it to vary in the interval $0.5\le\Delta_\sigma\lesssim0.8$. The lower end of this interval is fixed by the unitarity bound, while the upper end has been chosen arbitrarily. For each $\Delta_\sigma$ in this range, we ask: \emph{What is the maximal $\Delta_\en$ allowed by \reef{eq:cross}? }

\begin{figure}[htbp]
\begin{center}
\includegraphics[width=8cm]{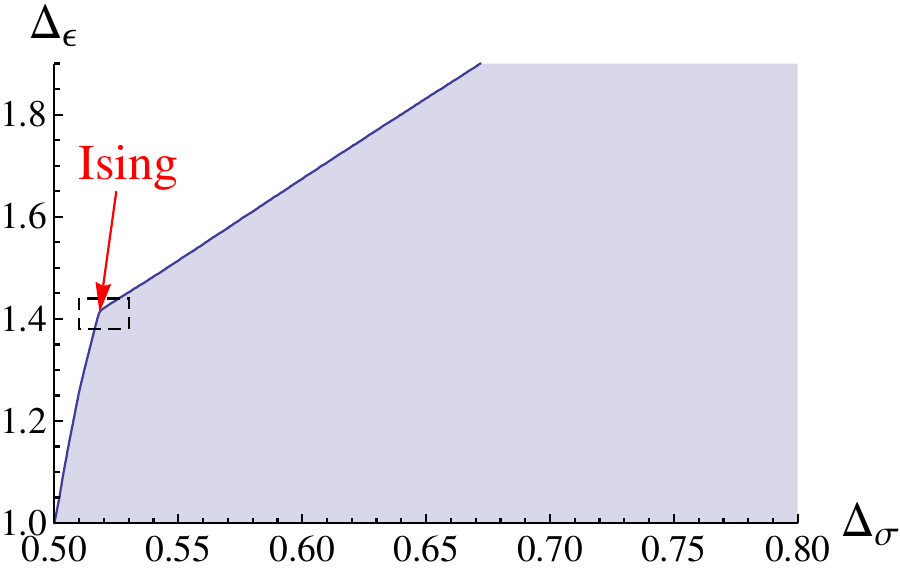}
\caption{\small Shaded: the part of the $(\Delta_\spin,\Delta_\en)$ plane allowed by the crossing symmetry constraint \reef{eq:cross}. The boundary of this region has a kink remarkably close to the known 3D Ising model operator dimensions (the tip of the arrow). The zoom of the dashed rectangle area is shown in Fig.~\ref{fig:deltaE-closeup}. This plot was obtained with the algorithm described in Appendix \ref{app:details} with $n_{\text{max}}=11$.}
\label{fig:deltaE}
\end{center}
\end{figure}
\begin{figure}[htbp]
\begin{center}
\includegraphics[width=8cm]{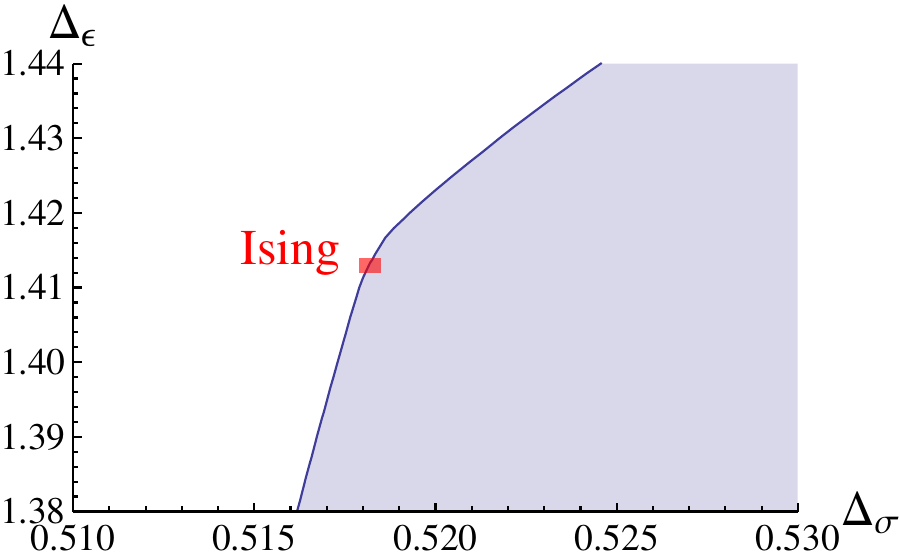}
\caption{\small The zoom of the dashed rectangle area from Fig.~\ref{fig:deltaE}. The small red rectangle is drawn using the $\Delta_\spin$ and $\Delta_\en$ error bands given in Table \ref{tab:dims}.}
\label{fig:deltaE-closeup}
\end{center}
\end{figure}

The result is plotted in Fig.~\ref{fig:deltaE}: only the points $(\Delta_\spin,\Delta_\en)$ in the shaded region are allowed.\footnote{To avoid possible confusion: we show only the upper boundary of the allowed region. $0.5\leq \Delta_\en\leq 1$ is also a priori allowed.} Just like similar plots in 4D and 2D \cite{Rattazzi:2008pe,Rychkov:2009ij,Poland:2011ey} the curve bounding the allowed region starts at the free theory point and rises steadily. Moreover, just like in 2D \cite{Rychkov:2009ij} the curve shows a kink whose position looks remarkably close to the Ising model point.\footnote{In contrast, the 4D dimension bounds do not show kinks, except in supersymmetric theories \cite{Poland:2011ey}.} This is better seen in Fig.~\ref{fig:deltaE-closeup} where we zoom in on the kink region. The boundary of the allowed region intersects the red rectangle drawn using the $\Delta_\spin$ and $\Delta_\en$ error bands given in Table \ref{tab:dims}. 

From this comparison, we can draw two solid conclusions. First of all, the old results for the allowed dimensions are not inconsistent with conformal invariance, though they are based on completely different techniques. Second, we can rigorously rule out about half of the $(\Delta_\spin,\Delta_\en)$ rectangle allowed by the table. It seems that the 3D Ising model lies remarkably close to the boundary of the allowed region, if not on the boundary. At present we don't have an explanation of why this had to be the case.

\subsection{Bounds Assuming a Gap Between $\en$ and $\en'$}

We will next give a series of plots showing the impact of assuming a gap in the $\bZ_2$-even scalar spectrum (as proposed in \cite{Rychkov:2011et}). In other words, we will impose that the first operator after $\en$ has dimension $\Delta_{\en'}$ above a certain value. 

Going from weaker to stronger, we will consider three constraints: $\Delta_{\en'}\ge 3,3.4,3.8$. Thus, we will ask: \emph{What is the region of the $(\Delta_\spin,\Delta_\en)$ plane allowed by \reef{eq:cross} when this extra constraint is taken into account?} 

\begin{figure}[htbp]
\begin{center}
\includegraphics[width=7cm]{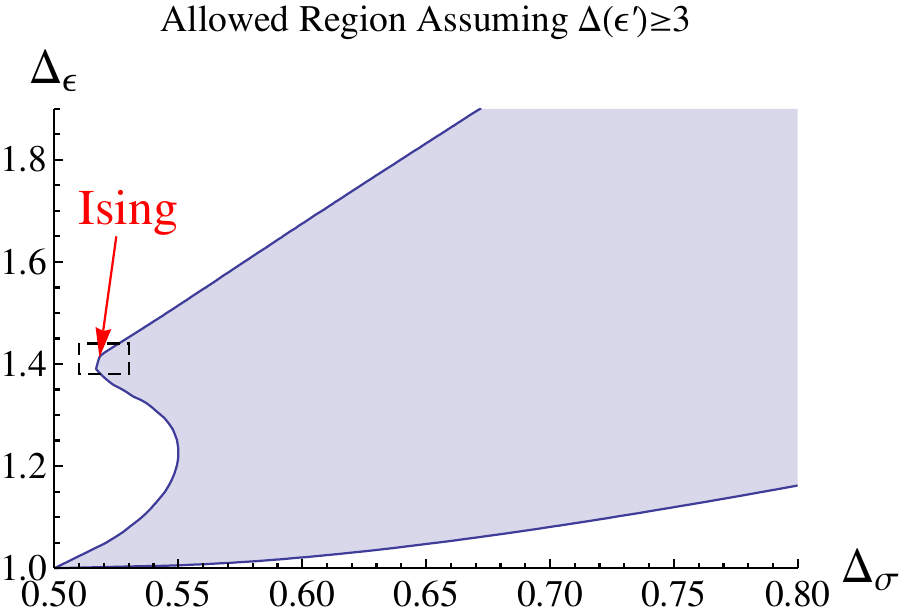}\hspace{1cm}
\includegraphics[width=7cm]{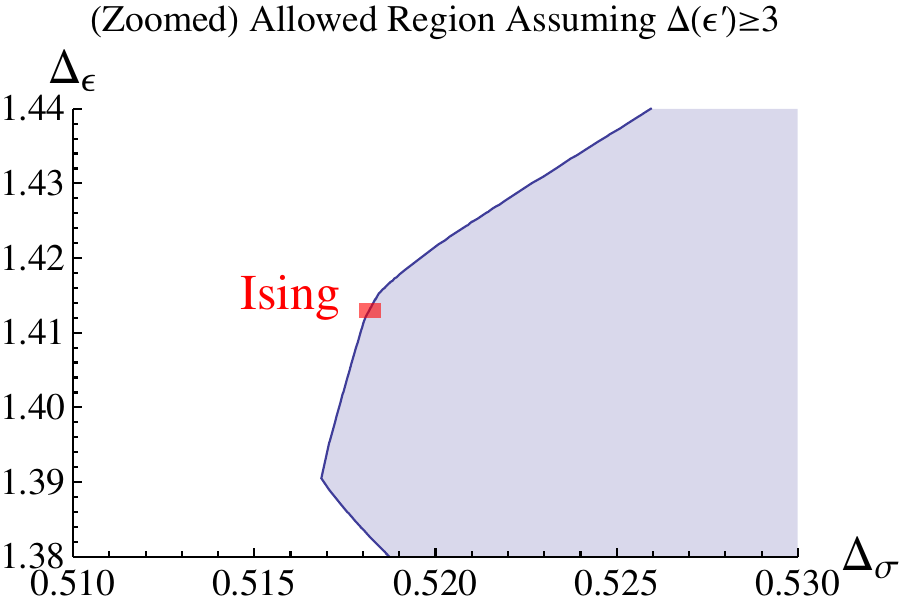}
\\(a)
\\[10pt]
\includegraphics[width=7cm]{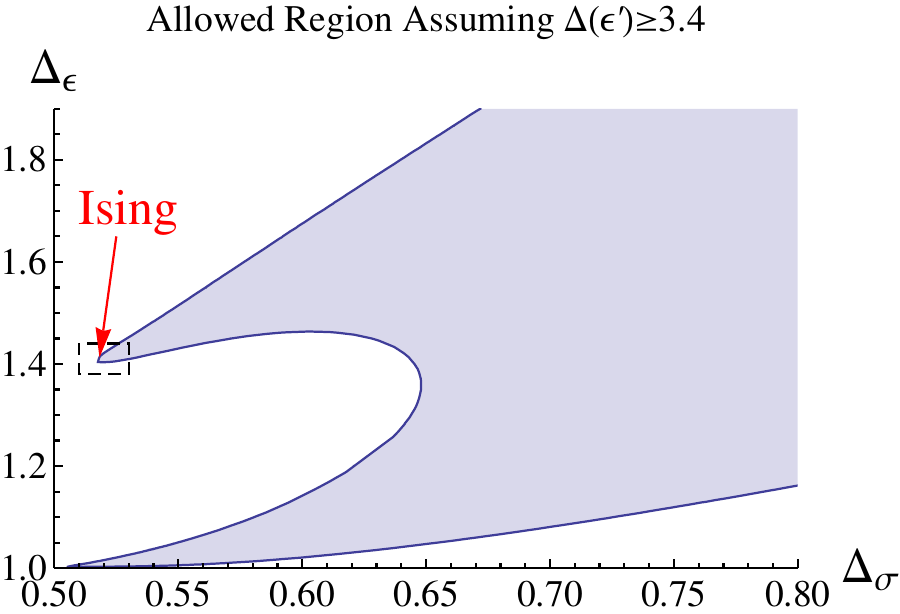}\hspace{1cm}
\includegraphics[width=7cm]{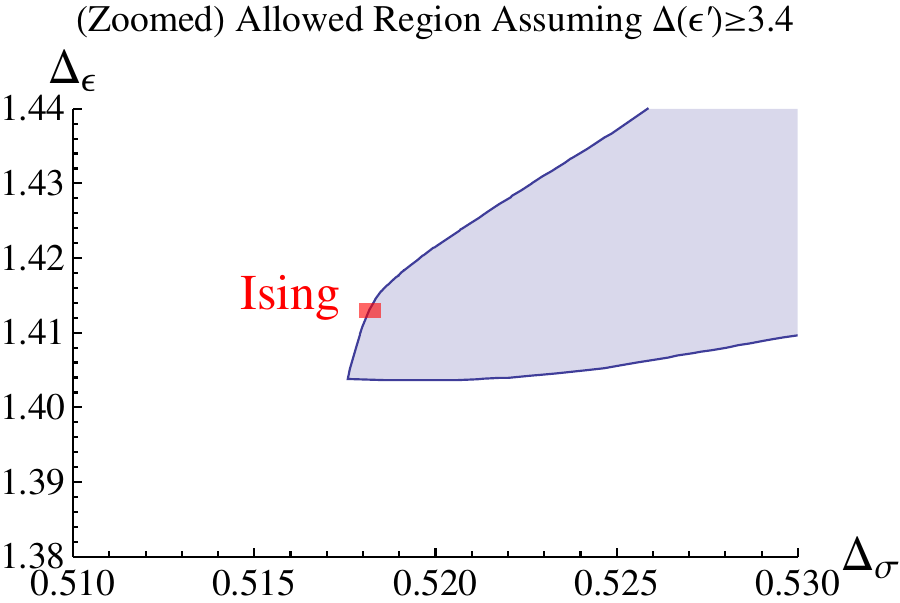}
\\(b)
\\[10pt]
\includegraphics[width=7cm]{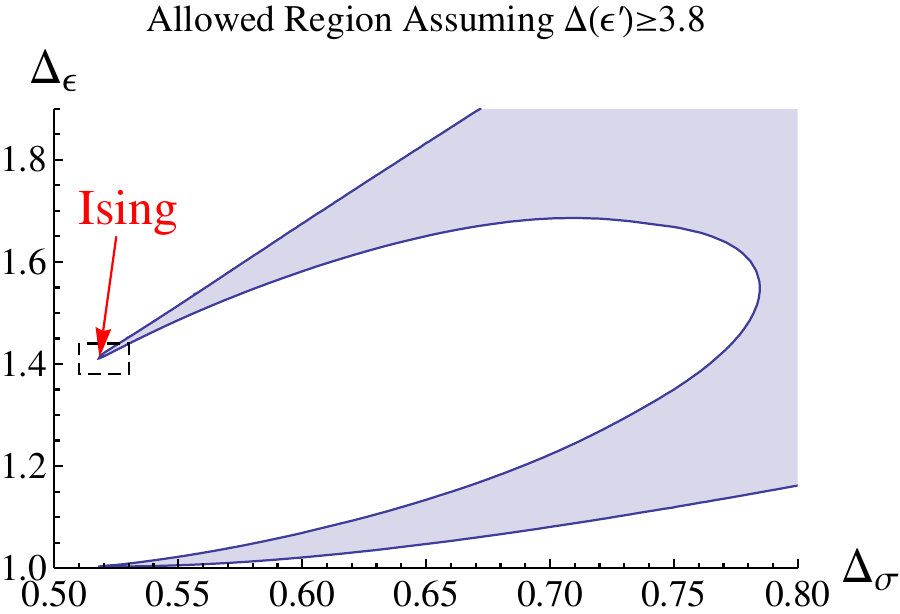}\hspace{1cm} 
\includegraphics[width=7cm]{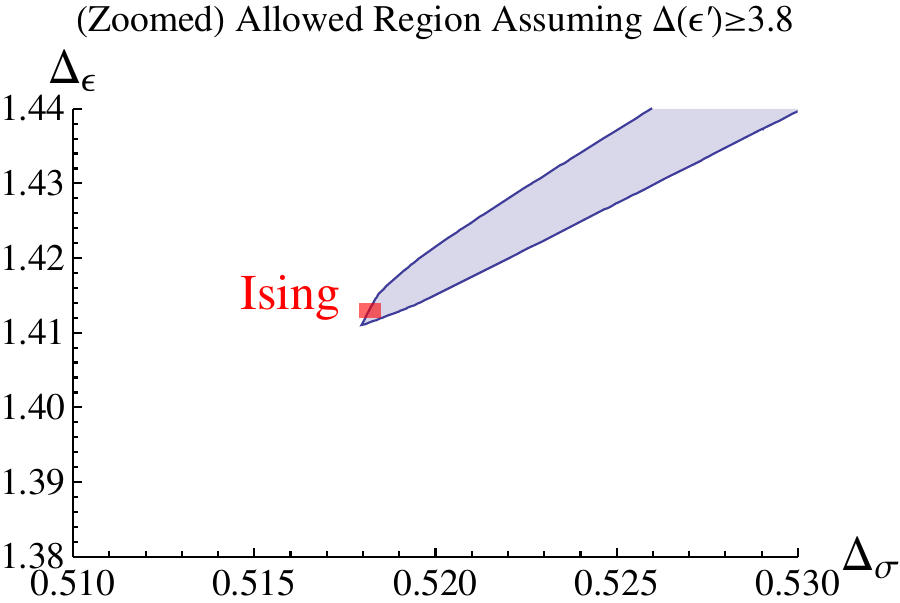}
\\(c)
\caption{\small Same as Figs.~\ref{fig:deltaE},~\ref{fig:deltaE-closeup}, but imposing the extra constraints $\Delta_{\en'}\ge\{3, 3.4, 3.8\}$.}
\label{fig:deltaE-Eprime-above}
\end{center}
\end{figure}

The weakest of the three assumptions, $\Delta_{\en'}\ge 3$, has been chosen since it can be justified experimentally: we know that the 3D Ising critical point is reached by fine-tuning just one parameter (the temperature). Therefore, it has just one relevant $\bZ_2$-even scalar, $\en$, while $\en',\en''$ etc.~must be irrelevant. As we see in Fig.~\ref{fig:deltaE-Eprime-above}(a), this piece of information allows to exclude a fair part of the region allowed by Fig.~\ref{fig:deltaE}. Unfortunately, close to the 3D Ising we do not gain constraining power: the new and the old bounds coincide there.

On the other hand, the stronger assumptions $\Delta_{\en'}\ge 3.4,3.8$ exclude a much larger portion of dimension space, carving out an allowed region with two branches; see Figs.~\ref{fig:deltaE-Eprime-above}(b,c).  The upper branch seems to end at the 3D Ising point, while the lower branch terminates near the free theory.  It is simple to understand why the intermediate region should not be allowed -- assuming a gap $\Delta_{\en'} > \Delta_*$ should exclude the gaussian line $\Delta_{\en} = 2\Delta_{\sigma}$ up to a dimension of $\Delta_{\sigma} = \Delta_*/2 - 1$, since the spectrum of this solution is $2\Delta_{\sigma} +2n + l$ for integer $n$. Our bounds are slightly weaker than that.

Zooming in on the tip near the 3D Ising point, we see that the allowed region in Fig.~\ref{fig:deltaE-Eprime-above}(c) barely intersects with the red rectangle. Were we to assume even larger gaps, the intersection would eventually disappear altogether. We performed this analysis and found that this happens for $\Delta_{\en'}\ge3.840(2)$. This result rules out the upper half of the $\Delta_{\en'}$ range allowed by Table \ref{tab:dims}, assuming that the more accurate determinations of $\Delta_\spin$ and $\Delta_\en$ in the same table are correct.

\begin{figure}[htbp]
\begin{center}
\includegraphics[width=8cm]{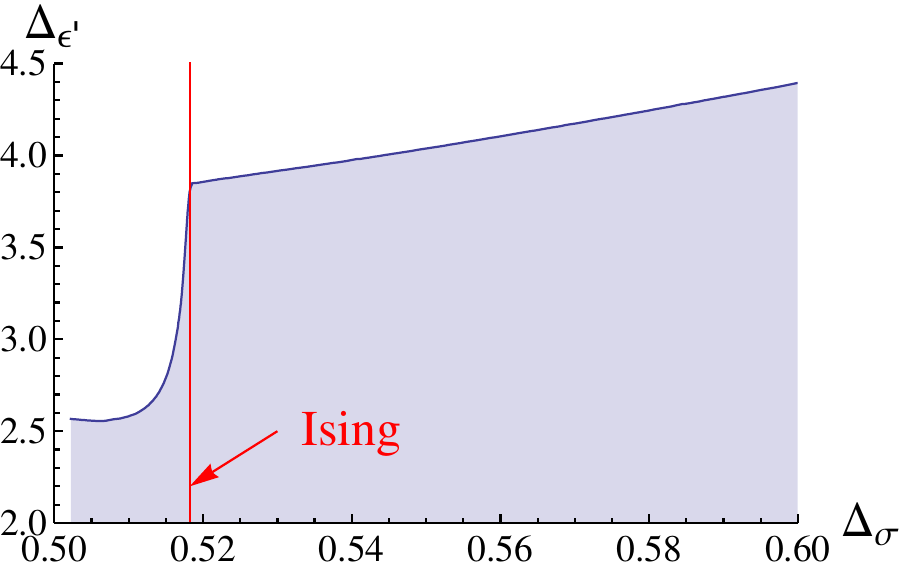}
\caption{\small The bound on $\Delta_{\en'}$ under the condition that $\Delta_{\en}$ has already been fixed to the maximal value allowed by Fig.~\ref{fig:deltaE}. Here $n_{\text{max}}=10$ (see Appendix \ref{app:details}). The width of the vertical red line marking the 3D Ising value of $\Delta_\sigma$ is about five times the error band in Table \ref{tab:dims}.}
\label{fig-epsprimebound}
\end{center}
\end{figure}

The same phenomenon is seen in a slightly different way in Fig.~\ref{fig-epsprimebound}. Here  
we compute the maximal allowed $\Delta_{\en'}$ under the condition that $\Delta_{\en}$ has already been fixed to the maximal value allowed by Fig.~\ref{fig:deltaE}. Notice the rapid growth of the $\Delta_{\en'}$ bound just below the 3D Ising model $\sigma$ dimension, which allows $\en'$ to become irrelevant. Similar growth has been observed in the 2D case in \cite{Rychkov:2011et}. Around the 3D Ising $\Delta_\sigma$ the bound is $\Delta_{\en'}\lesssim3.84$, consistent with the value cited above. 

This story illustrates how the conformal bootstrap equation imposes nontrivial dependencies between various operator dimensions. Once some dimensions are determined, the other ones are no longer arbitrary. Such interrelations are probably not easy to see from the renormalization group point of view. For instance, when using the $\eps$-expansion, each of the operator dimensions listed in Table \ref{tab:dims} requires an independent computation.

\subsection{Bounds on the Gap in the Spin 2 Sector}

The above discussion concerned the scalar sector of the 3D Ising model, but eventually we would like to also constrain operators with nonvanishing spin. For a first try, let's study here the gap in the spin 2 sector. The first spin 2 operator in the $\sigma\times\sigma$ OPE is the stress tensor $T_{\mu\nu}$, and we will be interested in the dimension of the second one, call it $T'_{\mu\nu}$. 

\begin{figure}[h]
\begin{center}
\includegraphics[width=8cm]{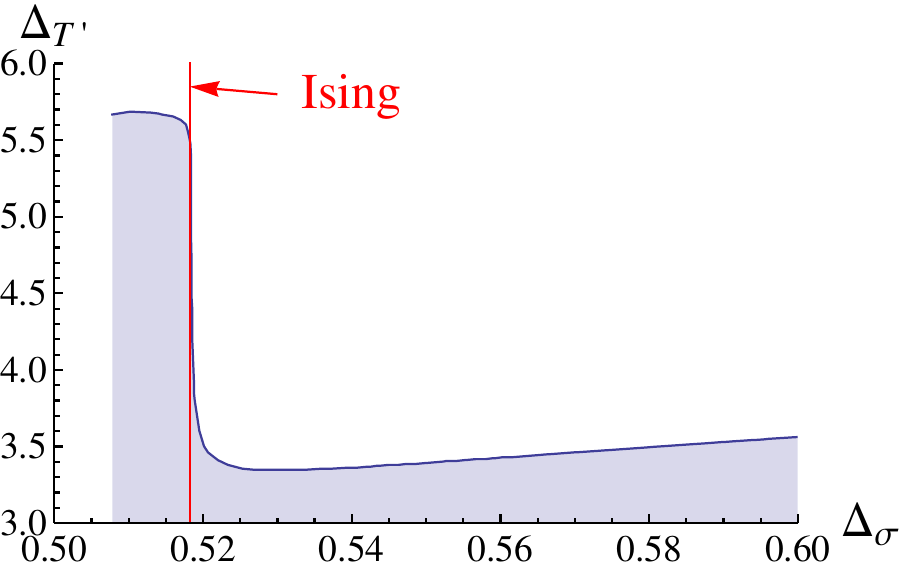}
\caption{Upper bound on the dimension of the second spin 2 operator $T'_{\mu\nu}$ from the crossing symmetry constraint \reef{eq:cross}. The algorithm from Appendix \ref{app:details} was used with $n_{\text{max}}=10$. The 3D Ising vertical red line is five times wider than the error band in Table \ref{tab:dims}. \SR{We do not show the region of $\Delta_\sigma$ close to the unitarity bound, which is subject to numerical instabilities.}}
\label{fig:Tprime}
\end{center}
\end{figure}

In Fig.~\ref{fig:Tprime} we give a rigorous upper bound on $\Delta_{T'}$ following from the crossing symmetry constraint \reef{eq:cross}. The bound is shown as a function of $\Delta_\sigma$ only, and is in this sense analogous to our first bound in Fig.~\ref{fig:deltaE}. Unlike for the case of $\en'$ studied in the previous section, we found that the bound on $T'$ is only very weakly correlated with the value of $\Delta_\en$, and so we do not show separately the allowed regions in the $(\Delta_\sigma,\Delta_{\en})$ plane. 

\SR{The $\Delta_{T'}$ bound shows fascinating behavior which is the opposite to that of Fig.~\ref{fig-epsprimebound}. It has a plateau at $\Delta_{T'}\sim5.7$ for low $\Delta_\sigma$ and suddenly drops to much lower values $\Delta_{T'}\sim3.5$ as the dimension of $\sigma$ is increased. To begin with, this implies that any moderate gap in the $T'$ dimension, e.g.~$\Delta_{T'}\ge4$, leads to a sharp upper bound on $\Delta_\sigma$. Taken together with the plots in Fig.~\ref{fig:deltaE-Eprime-above}, one then obtains very small \emph{closed} regions in the $(\Delta_\sigma, \Delta_\en)$ plane.}  

\SR{Moreover, the sudden drop in the $\Delta_{T'}$ bound happens precisely when $\Delta_\sigma$ passes the 3D Ising value.} The actual bound there is:
\beq
\Delta_\sigma \approx 0.518\quad\Longrightarrow\quad \Delta_{T'}\lesssim\SR{5.6}\,.
\label{eq:T'bound}
\eeq
Unfortunately, Table \ref{tab:dims} is mute about $\Delta_{T'}$ as we are not aware of any prior studies. However, we can get a rough estimate of this dimension by interpolating between 2D and 4D. In the 4D free scalar theory the first $\bZ_2$-even spin 2 operator after the stress tensor is
\beq
T'_{\mu\nu}=\,:\!\phi^2\, T_{\mu\nu}\!:\,\qquad\text{(4D)}\,,
\label{eq:4DT'}
\eeq
which has dimension 6. To be more precise, in the free scalar theory this operator is decoupled from the $\phi\times\phi$ OPE, but we expect it to couple in the Wilson-Fischer fixed point in $4-\eps$ dimensions.

In the 2D Ising model the first such \SR{quasiprimary operator is
\beq
T'=(L_{-4}-{\textstyle\frac{5}{3}} L_{-2}^2)\bar{L}_{-2} \mathbf{1}\qquad \text{(2D)}\,,
\label{eq:2DT'}
\eeq
again of dimension 6. 
Notice that another 2D candidate spin 2 quasiprimary, $(L_{-2}-\frac{3}{4} L_{-1}^2)\en$ of dimension 3, is a null state since the field $\en=\phi_{2,1}$ is degenerate on level 2 in the 2D Ising model.} 

Assuming as usual that the 2D Ising and the 4D free scalar are continuously connected by the line of Wilson-Fischer fixed points to which the 3D Ising model also belongs, we expect by interpolation that $\Delta_{T'}\approx 6$ in 3D, not far from the upper end of the range allowed by the rigorous bound \reef{eq:T'bound}.

\subsection{Bounds on Higher Spin Primaries}
\label{sec:higherspin}

In addition to bounding operators in the scalar and spin 2 sectors, we can also attempt to place bounds on higher spin primaries in the $\sigma \times \sigma$ OPE.  The first such operator in the 3D Ising model is the spin 4 operator $C_{\mu\nu\kappa\lambda}$. This operator is interesting because it controls the leading effects of rotational symmetry breaking when the 3D Ising model is placed on a cubic lattice. The corresponding perturbation of the CFT Lagrangian can be written as
\beq
\delta\mathcal{L}_{\text{CFT}}\propto C_{1111}+C_{2222}+C_{3333}\,.
\eeq
Because of this connection with phenomenology, the dimension of $C$ has been computed rather precisely: $\Delta_C \simeq 5.0208(12)$ (\cite{Campostrini:1999at}, Eq.~(4.9)).

\begin{figure}[h]
\begin{center}
\includegraphics[width=8cm]{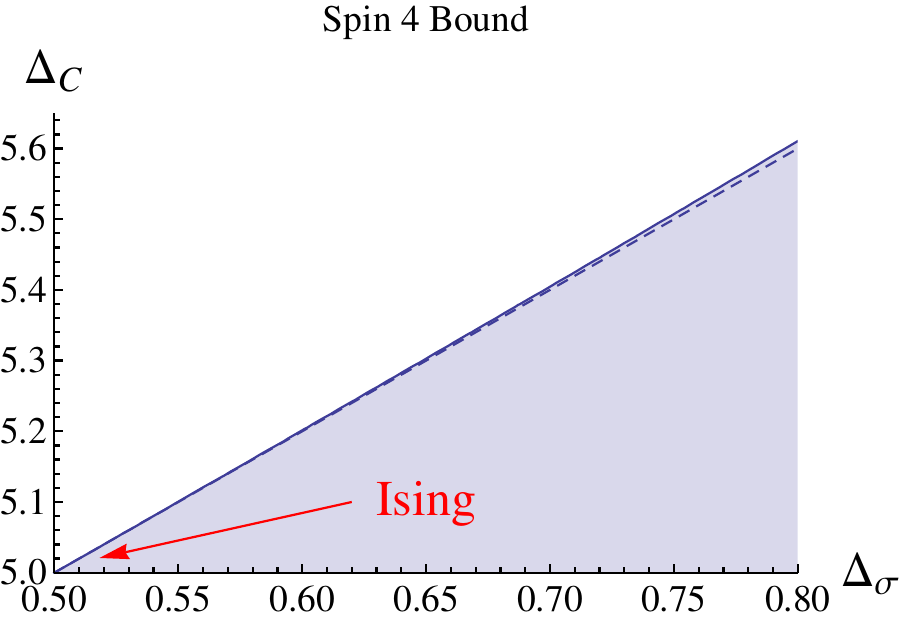}
\caption{Upper bound on the dimension of the first spin 4 operator in the $\sigma \times \sigma$ OPE from the crossing symmetry constraint \reef{eq:cross}. The algorithm from Appendix \ref{app:details} was used with $n_{\text{max}}=10$.  The tip of the arrow shows the point $(\Delta_\sigma,\Delta_C)$ with the 3D Ising model values from Table~\ref{tab:dims}. The dashed line is the gaussian solution $\Delta_4 = 2\Delta_{\sigma} + 4$. }
\label{fig:spin4}
\end{center}
\end{figure}

In Fig.~\ref{fig:spin4} we give a rigorous upper bound on $\Delta_C$ following from crossing symmetry and unitarity, making no other assumptions about the spectrum.  While this bound passes above the value of $\Delta_C$ in the 3D Ising model, this is easily understood by the fact that the gaussian solution to crossing symmetry has $\Delta_C = 2\Delta_{\sigma} + 4$, which must be respected by our bound.  The interesting and highly nontrivial statement is then that the gaussian solution seems to essentially saturate the bound.   The bound that we find is fit well by the curve:
\be
\Delta_C^{\textrm{max}} \simeq \left(2\Delta_{\sigma} + 4 \right) + 0.1176 \left(\Delta_{\sigma}-1/2 \right)^2 + O\left(\left(\Delta_{\sigma}-1/2 \right)^3 \right),
\ee
so that one can see that linear deviations are not required, and quadratic deviations are at least somewhat suppressed.  It is tempting to conjecture that the optimal bound (taking $n_{\textrm{max}} \rightarrow \infty$) will exactly follow the gaussian line.  It will be important in future studies to closely examine behavior of the bound at even larger external dimensions, to better understand whether deviations from this conjectured behavior are allowed.

\begin{figure}[h]
\begin{center}
\includegraphics[width=8cm]{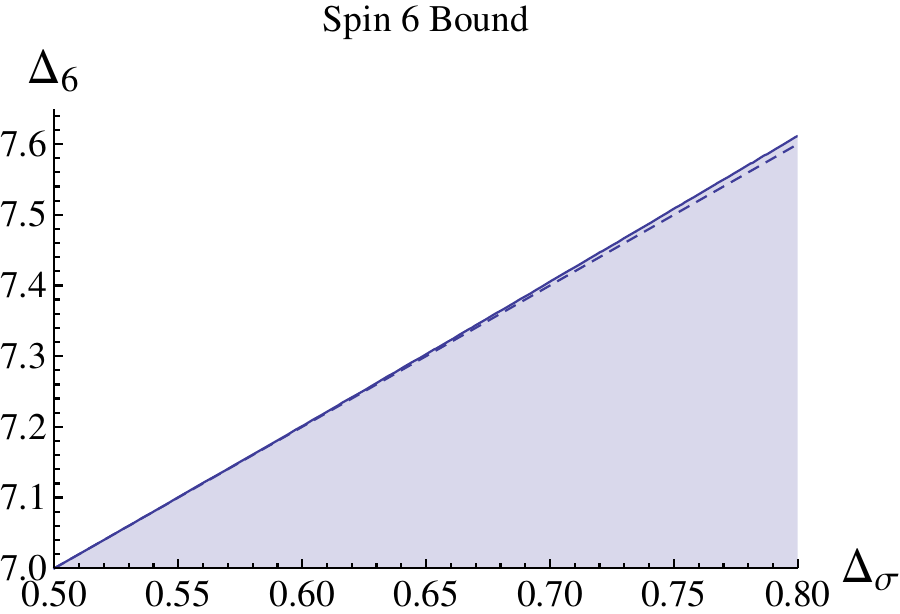}
\caption{Upper bound on the dimension of the first spin 6 operator in the $\sigma \times \sigma$ OPE from the crossing symmetry constraint \reef{eq:cross}. The algorithm from Appendix \ref{app:details} was used with $n_{\text{max}}=10$. The dashed line is the gaussian solution $\Delta_6 = 2\Delta_{\sigma} + 6$.}
\label{fig:spin6}
\end{center}
\end{figure}

Does this behavior of closely following the gaussian line hold for higher spins?  To explore this, in Fig.~\ref{fig:spin6} we show the analogous upper bound on the lowest-dimension spin 6 operator in the $\sigma \times \sigma$ OPE.  This operator would control breaking of rotational symmetry on the tetrahedral lattice, but we are not aware of prior 3D studies of its dimension. Again we see that the bound closely follows the gaussian line $\Delta_6= 2 \Delta_{\sigma} + 6$, with a fit:
\be
\Delta^{\textrm{max}}_6 \simeq \left(2\Delta_{\sigma} + 6 \right)  + 0.1307 \left(\Delta_{\sigma}-1/2 \right)^2 + O\left(\left(\Delta_{\sigma}-1/2 \right)^3 \right),
\ee
so that again quadratic deviations are suppressed. We have verified that this trend continues for operators of spin 8 and 10. 

An important feature of these bounds is that they approach the dimensions of spin $l$ conserved currents $\Delta_l = l+1$ as $\Delta_{\sigma} \rightarrow 1/2$.  It is well known that theories of free scalars contain higher spin conserved currents.  Our bound shows that theories containing almost-free scalars necessarily contain higher spin operators that are almost conserved currents.  A~CFT version of the Coleman-Mandula theorem proved recently in \cite{Maldacena:2011jn} shows that theories containing higher spin currents and a finite central charge necessarily have the correlation functions of free field operators. This implies that we should also be able to derive a {\it lower} bound on the dimensions of higher spin operators, perhaps under the assumption of a finite central charge.  It would be also interesting to connect these studies with an old result of Nachtmann \cite{Nachtmann:1973mr} that in a unitary theory the \emph{leading twists} 
\beq
\tau_l=\Delta_l-(l+D-2),
\eeq
where $\Delta_l$ is the dimension of the lowest spin $l$ operator, must form a \emph{nondecreasing} and \emph{convex upward} sequence for $l\ge2$. We leave exploration of these very interesting directions to future work.

These bounds are also particularly interesting in the context of the AdS/CFT correspondence, since they place tight constraints on $O(1/N^2)$ corrections to the dimensions of double-trace operators.  Concretely, free scalars in AdS give rise to spin-$l$ double-trace operators with gaussian dimensions $2\Delta_{\sigma} + 2n + l$ for integer $n$, while bulk interactions generate $O(1/N^2)$ corrections to these dimensions.  Some explicit examples of these corrections were studied, e.g., in~\cite{Heemskerk:2009pn,Fitzpatrick:2010zm}.  If our conjecture that the gaussian solution saturates the bound is true, then the bounds forbid bulk interactions that generate positive corrections to these dimensions, which in turn may imply positivity constraints on (higher derivative) interactions in AdS.  Such constraints could then be related to the constraints on higher derivative interactions studied in~\cite{Adams:2006sv}.  This is clearly another direction worth studying in future work.

\begin{figure}[h]
\begin{center}
\includegraphics[width=8cm]{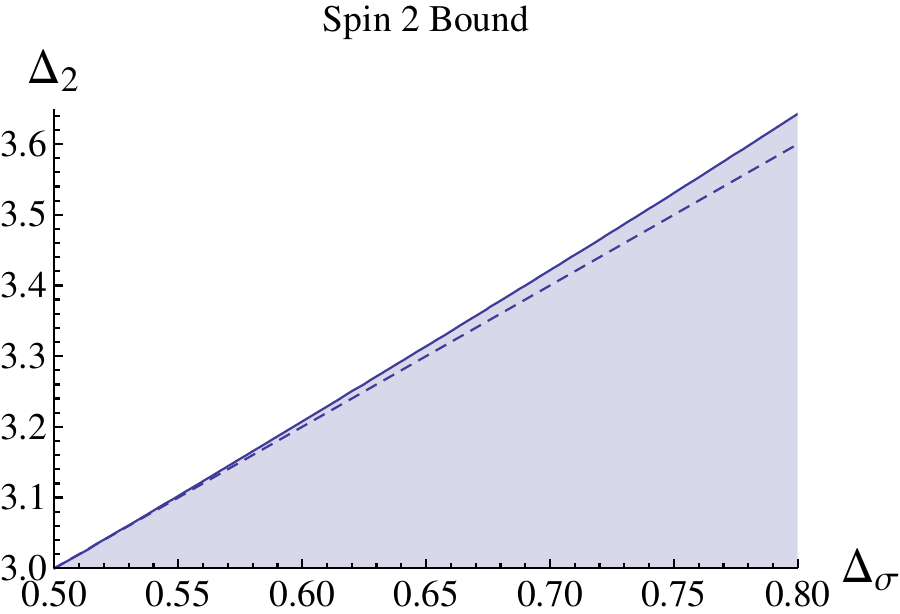}
\caption{Upper bound on the dimension of the first spin 2 operator in the $\sigma \times \sigma$ OPE from the crossing symmetry constraint \reef{eq:cross} in non-local theories without a stress tensor. The algorithm from Appendix \ref{app:details} was used with $n_{\text{max}}=10$. The dashed line is the gaussian solution $\Delta_2 = 2\Delta_{\sigma} + 2$.}
\label{fig:spin2}
\end{center}
\end{figure}

Finally, let us mention that similar bounds can be derived on the lowest dimension spin 2 operator in (non-local) theories where a stress tensor does not appear in the $\sigma \times \sigma$ OPE.  This bound (Fig.~\ref{fig:spin2}) shows similar features to the higher spin bounds.  

Such non-local theories may be interesting for several reasons. First, they commonly arise in statistical mechanics as models of long-range critical behavior. One much studied example is the critical point of the \emph{long-range Ising model}, defined by a lattice Hamiltonian with a power-law spin-spin interaction:
\beq
\mathcal{H}=-\sum_{i,j}\frac{s_i s_j}{r_{ij}^{D+\gamma}}\,.
\label{eq:lr}
\eeq
The precise universality class of this model depends on the value of $\gamma$. According to classic results \cite{Fisher:1972zz,Sak} supported by Monte-Carlo simulations \cite{Luijten}\footnote{Note added: very recent accurate Monte-Carlo simulations \cite{Picco} indicate deviations near the boundary of the intermediate and the short-range regions from the established picture as described below.} , there are three regions. For $\gamma$ sufficiently small, namely $\gamma\le D/2$, the critical point is the gaussian model with the spin-field dimension determined by the na\"ive continuous limit of \reef{eq:lr}: $\Delta_\sigma=(D-\gamma)/2$. Then there is an intermediate region, and finally the region of large $\gamma$, in which the model
belongs to the usual, short-range, Ising model universality class and the critical exponents do not depend on $\gamma$. The boundary between the intermediate and short-range region lies at $\gamma=D-2\Delta^\text{Ising}_\sigma$, determined by the short-range Ising model spin-field dimension. This can be also understood by studying stability of the short-range Ising model with respect to non-local perturbations. Analogously, the boundary between the gaussian and the intermediate region lies at the value of $\gamma$ for which the operator $\sigma^4$ becomes marginal. 

In the intermediate region, the $\sigma$ dimension is still given by the mean-field formula $\Delta_\sigma=(D-\gamma)/2$, but the dimensions of other operators, such as $\en$, have nontrivial dependences on $\gamma$ deviating from the gaussian values. So these fixed points are interacting. Because of their origin as relevant perturbations of the non-local gaussian scalar theory, they are expected to have conformal symmetry (and not just scale invariance), but not a stress-tensor.
It is for such non-local CFTs that our bound in Fig.~\ref{fig:spin2} may be of interest.

Another reason to be interested in theories without a stress tensor is that they realize a simpler case of AdS/CFT, in which bulk gravity is decoupled, so that the AdS metric is viewed as a fixed non-fluctuating background.\footnote{Such theories may alternately be viewed as the starting point for ``constructive holography'' by defining a CFT perturbatively around the $c\rightarrow \infty$ point as done in e.g. \cite{El-Showk2011a}.}  This may be useful when one is interested in aspects of the correspondence which are not necessarily related to gravity, as e.g.~in \cite{Heemskerk:2009pn}. Also, removing gravity allows one to find nontrivial UV-complete AdS/CFT examples which are purely field-theoretic (no strings): any UV-complete quantum field theory on the AdS${}_{D+1}$ background can be interpreted as providing a dual description to a non-local $D$-dimensional CFT on the boundary.

\subsection{Bounds on the Central Charge}

Our final application concerns the central charge $C_T$ of the 3D Ising model, defined for an arbitrary $D$ as the coefficient of the canonically normalized stress tensor two-point function:
\begin{align}
\left\langle T_{\mu\nu}(x)T_{\lambda\sigma}(0)\right\rangle  &  =\frac{C_{T}%
}{S_{D}^{2}}\frac{1}{(x^{2})^{D}}\left[  \frac{1}{2}(I_{\mu\lambda}%
I_{\nu\sigma}+I_{\mu\sigma}I_{\nu\lambda})-\frac{1}{D}\delta_{\mu\nu}%
\delta_{\lambda\sigma}\right]  \,,\nonumber\\
I_{\mu\nu}  &  =\delta_{\mu\nu}-2x_{\mu}x_{\nu}/x^{2}\,,\qquad S_{D} = 2\pi^{D/2}/\Gamma(D/2)\,. \label{eq:TT}%
\end{align}
It seems that the 3D Ising central charge has been computed only to the second order in the $\eps$-expansion, with the result \cite{Hathrell:1981zb,Jack:1983sk,Cappelli:1990yc,Petkou:1994ad}
\beq
C_T/C_T^{\text{free}}=1-\frac5{324} \eps^2+O(\eps^3)\,,
\eeq
where $C_T^{\text{free}}=D/(D-1)$ is the free scalar field central charge. Substituting $\eps\to 1$ and neglecting the unknown higher-order terms, this estimate would suggest that $C_T/C_T^{\text{free}}$ is very close to 1, around 0.98 or so.

\begin{figure}[htbp]
\begin{center}
\includegraphics[width=8cm]{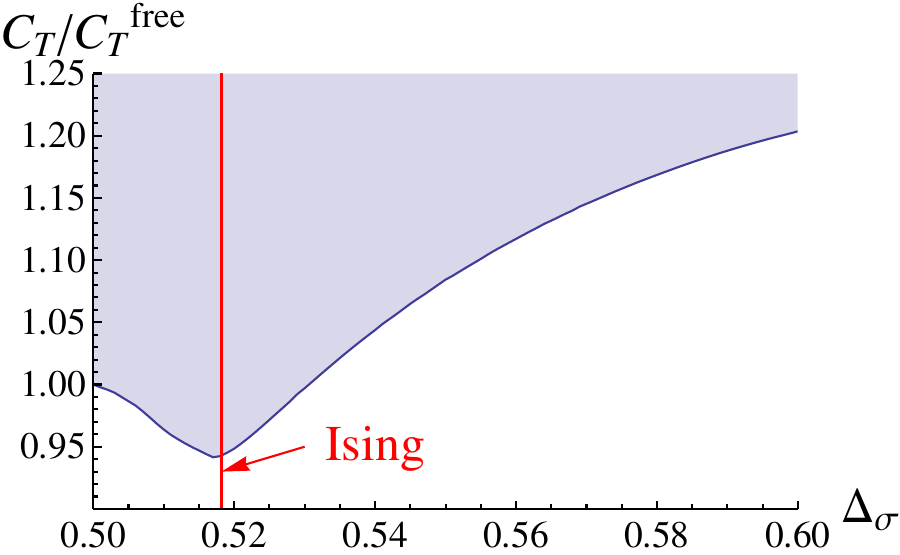}
\caption{\small The lower bound on $C_T$ as a function of $\Delta_\sigma$. The plot was obtained with $n_{\text{max}}=11$. The 3D Ising vertical red line is five times wider than the error band in Table \ref{tab:dims}.}
\label{fig:central-charge}
\end{center}
\end{figure}

In our method, we can get control over $C_T$ because the stress tensor conformal block enters the crossing symmetry constraint \reef{eq:cross} with a $C_T$ dependent coefficient:\footnote{The prefactor is different from \cite{Rattazzi:2010gj} due to the different conformal block normalization, see Eq.~\reef{eq:fac}.}
\beq
p_{D,2}=\frac{D}{D-1} \frac{\Delta_\spin^2}{C_{T}}\,.
\label{eq:p42}
\eeq
Following \cite{Poland:2010wg,Rattazzi:2010gj}, the conformal bootstrap can be used to bound the coefficient $p_{3,2}$ from above, which bounds the central charge from below. In Fig.~\ref{fig:central-charge} we show the lower bound on $C_T$ as a function of $\Delta_\sigma$. We see that the bound has a distinctive minimum close to the 3D Ising value of the $\spin$ dimension. The position of the minimum corresponds to $C_T/C_T^{\text{free}}\approx 0.94$. One may also redo the plot in Fig.~\ref{fig:central-charge} making some assumption about $\Delta_\en$, like that $\Delta_\en\ge\Delta_\spin$. The most aggressive assumption would be to fix $\Delta_\en$ to the maximal value allowed by the upper bound in Fig.~\ref{fig:deltaE}. One finds that the shape of the bound on $C_T$ is very weakly dependent on these assumptions, but that the minimum moves to the right, even closer to the 3D Ising $\Delta_\sigma$, and slightly higher up to $C_T/C_T^{\text{free}}\approx 0.95$.

We believe that the observed minimum in the $C_T$ lower bound is not accidental, but must be close to the true value of $C_T$.\footnote{In 2D, a similar analysis reproduces the exact value of the 2D Ising model central charge with $10^{-4}$ accuracy \cite{Vichi-thesis}.}
This would imply a small but noticeable discrepancy with the $\eps$-expansion estimate of $C_T$, which can be attributed to the unknown higher-order terms. In fact, we can also derive \emph{upper} bounds on $C_T$ in presence of a gap between $T$ and $T'$. The strength of these bounds depends on the assumption about the gap, and for $T'$ close to the maximal value allowed by \reef{eq:T'bound} would rigorously rule out the $\eps$-expansion estimate. We leave full exploration of such upper bound bounds to future work.

\section{Discussion}
\label{sec:discussion}

The results of the previous section have many implications whose importance is hard to overestimate.  First, all of our bounds are consistent with everything that was previously known about the critical exponents of the 3D Ising model, as computed via RG methods and measured in experiments and Monte Carlo simulations.  We should take this a very strong evidence that the 3D Ising model has a full conformal symmetry, justifying post factum the use of conformal symmetry in studying this theory.  It would be good to further test the conformal invariance experimentally or on the lattice, for example by measuring the form of the 3-point functions.  One can also compare any new measurements (e.g.,~of the central charge) against the constraints obtained using the methods in this paper.

It is worth emphasizing that the bootstrap approach to studying 3D CFTs taken in this paper has, in principle, a significant advantage over other methods -- at every step in the program we can present constraints that are completely rigorous (up to numerical errors that can be made arbitrarily small).  This is significantly better than the usual situation in field theory computations, where one computes the first several terms in a series (say, the $\eps$-expansion or a loop expansion) and one can only estimate the errors from neglecting higher terms.  It is also an advantage over lattice simulations, where it can be very difficult to gain control over errors induced by discretizing the theory.

In this paper we have imposed only the first and the simplest of the infinitely many bootstrap conditions -- the one following from the crossing symmetry of the $\sigma$ four-point function. It turns out that this condition alone carves out a significant portion of the operator dimension space. The 3D Ising model seems to lie on the boundary of the allowed region, and at a rather special point -- a corner. This empirical fact suggests that the model is algebraically special, for two reasons. 
First, the crossing symmetry constraint is expected to allow fewer solutions at the boundary of the allowed region as compared to the bulk, perhaps just a unique solution. Second, the non-analytic behavior of the bound at a corner point can be attributed to rapid rearrangements of the operator spectrum \cite{Rychkov:2011et}. Indeed, Figs.~\ref{fig-epsprimebound} and \ref{fig:Tprime} show rapid changes happening for the next-to-leading operator dimensions in the scalar and spin 2 sectors. 
Such spectrum rearrangements signal linear (near-)degeneracies among various conformal blocks.
It is very important to explore this phenomenon in detail as it offers tantalizing hope for distilling some \emph{analytical} understanding of the 3D Ising model dimensions from our numerical approach.  More generally, the fact that some special theories seem to lie at the edge of the region allowed by crossing symmetry may suggest a new classification scheme for understanding CFTs in $D>2$.

Furthermore, it is intriguing that most of our bounds (not just $\Delta_\en$) seem to be essentially saturated by the values realized in the 3D Ising model.  This fact suggests the strategy of determining the spectrum recursively: first fix $\en$ at the maximal allowed dimension, then $\en'$ at the maximal allowed dimension given $\Delta_\en$, etc. We hope to explore the viability of this approach (perhaps also including gaps in higher spin operators) in future work.  

Another badly needed development is to add conformal bootstrap constraints coming from other correlators, which can lead to interesting interplay. 
 For example, we would like to include $\langle\sigma \en \sigma \en \rangle$ expanding in the $\sigma\times\en$ channel, since this expansion will be crossing symmetric. Moreover, the conformal block of $\sigma$ will appear with the same coefficient $f^2_{\sigma\sigma\en}$ as the conformal block of $\en$ in the analysis of $\langle\sigma \sigma \sigma \sigma \rangle$. It is also interesting to include $\langle\en \en \en \en \rangle$ whose expansion involves the same $\bZ_2$-even operators as $\langle\sigma \sigma \sigma \sigma \rangle$.  Due to \reef{eq:p42}, the stress tensor will appear in both expansions with related coefficients.  One can also consider 4-point functions containing the stress tensor, where the recent results of~\cite{Costa:2011dw} on conformal blocks for external operators with spin can be used. 
 
Another future task is to study 3D CFTs with larger global symmetry groups, such as $O(N)$ symmetry. The general theory of analyzing bootstrap constraints in the presence of a continuous global symmetry was given in \cite{Rattazzi:2010yc}. The equations look more difficult as the OPE contributions should be classified into various representations and the crossing symmetry transformation involves a Fierz matrix. Nevertheless, there are always as many equations as representation channels and the total constraining power is expected to be comparable to the $\bZ_2$-symmetric case. In 4D, this has been convincingly demonstrated in \cite{Vichi:2011ux,Poland:2011ey}, where many strong bounds for $O(N)$ and $SU(N)$ symmetric CFTs have been obtained.  It would be interesting to generalize these methods to 3D and see how the resulting bounds on operator dimensions compare to what is known about the $O(N)$-vector models.  

Cross-fertilizing in the opposite direction, it is worth applying in 4D what we have learned in this paper in the 3D context -- how interesting it is to study the effects of gaps in the operator spectrum. In addition, we should stress that the recursion relations for conformal blocks exhibited in this paper are valid for any space-time dimension $D$.  Thus, we can use them to numerically compute conformal blocks in $4-\eps$ dimensions for different values of $\epsilon$, where we can make contact with operator dimensions and OPE coefficients computed perturbatively in the $\eps$-expansion.

Our results and discussion in Section~\ref{sec:higherspin} show that one can also learn interesting statements about higher spin operators from crossing symmetry.  It will be interesting to explore what can be learned further, particularly in the context of the AdS/CFT correspondence, where for example the $O(N)$-vector models in the large $N$ limit are described by higher spin gauge theories in AdS${}_4$~\cite{Klebanov:2002ja,Sezgin:2002rt}.  There is clearly still much to be learned about the role that higher spin operators play in ensuring consistency of the theory, and about how gaps in the lower-spin spectrum affect what these operators are allowed to do.

Overall, our results fly in the face of the prevailing opinion that above two dimensions conformal symmetry by itself is not sufficiently restrictive to solve models.  Clearly, the conformal bootstrap in $D > 2$ works.  We have not yet solved the 3D Ising model, but we have definitely cornered it.

\section*{Acknowledgements}
We are grateful to Zohar Komargodski, Juan Maldacena, Hugh Osborn, Mohammad Rajabpour and Sasha Zhiboedov for useful discussions and remarks. The work of S.R. is supported in part by the European Program ÒUnification in the LHC EraÓ, contract PITN-GA-2009-237920 (UNILHC), and by the \'Emergence-UPMC-2011 research program.  The work of D.P. is supported by D.O.E. grant DE-FG02-90ER40542.  The work of A.V. is supported by the Office of High Energy Physics of the U.S. Department of Energy under the Contract DE-AC02-05CH1123.  The work of S.E. is supported primarily by the Netherlands Organization for Scientific Research (NWO) under a Rubicon grant and also partially by the ERC Starting Independent Researcher Grant 240210 - String-QCD-BH.  

The computations in this paper were run on several different clusters: the Odyssey cluster supported by the FAS Science Division Research Computing Group at Harvard University, the Aurora cluster supported by the School of Natural Sciences Computing Staff at the Institute for Advanced Study, and the Kelvin cluster at the C.E.A. Saclay funded by the European Research Council Advanced Investigator Grant ERC--AdG--228301.  S.E. would like to thank D. Kosower for providing access to the Kelvin cluster.

\newpage
\appendix
\section{Recursion Relations at Fixed External Dimensions}
\label{app:recursion}

Our conformal blocks are the same as the functions $F_{\lambda_1,\lambda_2}$ of Ref.~\cite{DO3}:
\beq
G_{\Delta,l}=F_{\lambda_1,\lambda_2},\qquad\lambda_{1}=\half(\Delta+l),\ \lambda_{2}=\half(\Delta-l)\,.
\eeq
This normalization is different from the one used in a number of previous works. E.g.,~in \cite{DO1,Costa:2011dw} conformal blocks contain an extra factor of\footnote{$\alpha$ was called $\eps$ in \cite{DO3}, but we already have two other epsilons in this paper.}
\beq
\frac{(2\alpha)_l}{(-2)^l (\alpha)_l}\,,\qquad \alpha\equiv \frac D2-1\,.
\label{eq:fac}
\eeq
This follows by comparing Eq.~(2.25-29) of \cite{DO3} with Eq.~(2.22) of \cite{Costa:2011dw}. 

We note in passing one reason for using the new normalization: once conformal blocks are analytically continued to all real $l$, one has a symmetry relation (\cite{DO3}, Eq.~(4.10))
\beq
G_{\Delta,l}=G_{\Delta,-l-D+2}\,.
\label{eq:mirror}
\eeq
In particular, we have $G_{\Delta,-1}=G_{\Delta,0}$ in 3D, which can be useful as explained in footnote \ref{note:-1}.

Below we consider only the case $\Delta_{12}=\Delta_{34}=0$, which corresponds to setting $a=b=0$ in the notation of \cite{DO3}. Denote
	\begin{gather}
	\beta_p\equiv \frac{p^2}{4(2p-1)(2p+1)}\,, \qquad
	D_z\equiv z^2(1-z) \partial_z^2-z^2 \del_z\,, \\
	\mathcal F_0\equiv \frac{1}{z}+\frac{1}{\bar z}-1\,,\qquad
	\mathcal F_1\equiv (1-z)\del_z+(1-\bar z) \del_{\bar z}\,,\qquad
	\mathcal F_2\equiv \frac{z-\bar z}{z \bar z}\left(D_z-D_{\bar z}\right)\,.
	\end{gather}
It was shown in \cite{DO3} that $\calF_i  \, F_{\lambda_1\lambda_2}$ can be expressed as linear combinations of 
$F_{\lambda'_1\lambda'_2}$. More specifically, we have (see \cite{DO3}, Eqs.~(4.28), (4.29), (4.32))
\begin{align}
\calF_0  \, F_{\lambda_1\lambda_2} &= 
{{l+2\alpha}\over {l+ \alpha}} \, F_{\lambda_1\lambda_2{-1}} +
 {l \over l +  \alpha }\, F_{\lambda_1{-1}\lambda_2} \nn\\
&{}+{ (\Delta-1)(\Delta -2\alpha)\over
(\Delta -1 -  \alpha)(\Delta-  \alpha )}
\bigg (
{l + 2\alpha \over l + \alpha } \, \beta _{\lambda_1}  
F_{\lambda_1{+1}\lambda_2} +  {l \over l +  \alpha } \, 
\beta_{\lambda_2 - \alpha} F_{\lambda_1\lambda_2{+1}}\bigg )\, ,
\label{eq:F0}\\
\calF_1  \, F_{\lambda_1\lambda_2} &= \frac{l+2 \alpha }{l+\alpha }\lambda _2\, F_{\lambda
   _1\lambda _2-1}
   +\frac{l
  }{l+\alpha }  \left(\lambda _1+\alpha \right) F_{\lambda _1-1\lambda
   _2}+\frac{(\Delta -1) (\Delta -2 \alpha ) }{(\Delta -1-\alpha ) (\Delta -\alpha )}\times \nn\\
   &
   \times\left( 
   \frac {l+2
   \alpha   }{l+\alpha }
  \left(-\lambda _1+\alpha +1\right) \beta _{\lambda _1}  F_{\lambda _1+1\lambda
   _2}
  +\frac{l }{ l+\alpha}
   \left(-\lambda _2+2 \alpha
   +1\right) \beta _{\lambda _2-\alpha}
F_{\lambda _1\lambda _2+1}     
 \right)\,, 
 \label{eq:F1}\\
\calF_2 \,  F_{\lambda_1\lambda_2} &=  (\Delta - 1 ) \, {l(l+2\alpha)\over l+\alpha}  
\bigg [ F_{\lambda_1\lambda_2{-1}} - F_{\lambda_1{-1}\lambda_2}
\nn\\
& \hskip 0.5cm
-{(\Delta- 2\alpha )(\Delta -1 -2\alpha)\over
(\Delta -1 -  \alpha)(\Delta-  \alpha )}\,
\big ( \beta _{\lambda_1}\,
F_{\lambda_1{+1}\lambda_2} - \beta_{\lambda_2 - \alpha} \, 
F_{\lambda_1\lambda_2{+1}} \big )\bigg ]\,,
 \label{eq:F2}\end{align}
where $\Delta= \lambda_1+\lambda_2$, $l=\lambda_1-\lambda_2$.

Let us now view \reef{eq:F0} and \reef{eq:F2} as a linear $2\times 2$ system for the spin $l+1$ conformal blocks $F_{\lambda_1+1\lambda_2}$ and $F_{\lambda_1\lambda_2-1}$. Eliminating one of these, say $F_{\lambda_1+1\lambda_2}$, we get a recursion relation expressing the remaining spin $l+1$ block in terms of spin $l$ and spin $l-1$ blocks only. Shifting the spin by one and passing to the $G_{\Delta,l}$ notation, this relation takes the form: 
\begin{multline}
\frac{(\Delta -\alpha ) (l+2
   \alpha -1)}{l+\alpha -1} G_{\Delta ,l}=
\frac{\alpha  (\Delta +l-1)}{l+\alpha -1} G_{\Delta ,l-2}+\frac{1}{2} \left((\Delta -2 \alpha )
   \calF_0+\frac{\calF_2}{
   l-1}\right)G_{\Delta +1,l-1}\\
   -\frac{\Delta  (\Delta -2 \alpha ) (\Delta -2 \alpha
   +1)  }{(\Delta -\alpha ) (\Delta -\alpha
   +1)} \beta_{\frac{1}{2} (\Delta
   -l+2-2\alpha)}G_{\Delta +2,l-2}\,. 
   \label{eq:rec1}
   \end{multline}
When we specialize to the line $z=\bar{z}$, the term involving $\calF_2$ vanishes. We are then left with a nonderivative recursion relation, Eq.~\reef{eq:recz=zb} of the main text.

Alternatively, we can apply the same logic to the system formed by \reef{eq:F0} and \reef{eq:F1}. Eliminating again $F_{\lambda_1+1\lambda_2}$ in favor of $F_{\lambda_1\lambda_2-1}$, shifting the spin by one and passing to the $G_{\Delta,l}$ notation, we get: 
\begin{multline}
\frac{(\Delta -\alpha ) (l+2
   \alpha -1)}{(l+\alpha -1)} G_{\Delta ,l}=
\left(\frac{1}{2}(\Delta +l-2
   \alpha -2) \calF_0 +\calF_1\right)G_{\Delta +1,l-1} 
   \\-(l-1)\left(\frac{\Delta  (\Delta
  -2 \alpha +1) }{(\Delta -\alpha )
   (\Delta -\alpha +1)}\beta_{\frac{1}{2} (\Delta
   -l+2-2\alpha)}G_{\Delta +2,l-2}
    +\frac{\Delta +l-1 }{l+\alpha -1}G_{\Delta
   ,l-2}\right)\,.
   \label{eq:rec2}
    \end{multline}
    
 Recursions \reef{eq:rec1} and \reef{eq:rec2} have complementary advantages. The first one becomes nonderivative at $z=\bar z$ and can be used to compute high spin blocks on this line efficiently. However, it needs both $l=0$ and $l=1$ blocks to start up (except in $D=3$ where it can be started from $l=0$ and $l=-1$, but we would like a framework which works in any $D$). On the other hand, recursion \reef{eq:rec2} has spin $l-2$ blocks entering with a factor $(l-1)$ and can be started up with just spin 0. In Appendix \ref{app:scalarblocks}, we'll use~\reef{eq:rec2} to compute spin 1 blocks at $z=\bar z$ from spin 0, but switch to the nonderivative recursion \reef{eq:rec1} for higher spins.

\section{Scalar and Spin 1 Blocks at $z=\bar z$}\label{app:scalarblocks}

In this appendix we'll derive formulas for the spin 0 and 1 conformal blocks at $z=\bar z$ for equal external dimensions.

We start with the double series expansion \reef{eq:doublesum} for the scalar conformal block. Performing the summation in $n$, we get ($\alpha=D/2-1$)
   \beq
  G_{\Delta,0}=\sum_{m=0}^\infty \frac{\left(\left(\frac{\Delta }{2}\right)_m\right){}^4  }{m! (\Delta )_{2 m} (\Delta -\alpha )_m}
   u^{\frac{\Delta
   }{2}+m}\, _2F_1\left(m+\frac{\Delta }{2},m+\frac{\Delta }{2};2 m+\Delta
   ;1-v\right)\,.
   \eeq
We now replace ${}_2F_1$ by its Euler integral representation 
\beq
	_2 F_1(a,b;c;x)=\frac{\Gamma(c)}{\Gamma(b)\Gamma(c-b)} \int_0^1 dt\, \frac{t^{b-1} (1-t)^{c-b-1}}{(1-t x)^a} \,.
	\label{eq:eulerint}
\eeq
The series in $m$ under the integral sign turns out to be hypergeometric in the variable
\beq
X=\frac{(1-t) t u}{1-t(1-v)}\,,
\eeq
so that we find:
\beq
G_{\Delta,0}=\frac{\Gamma (\Delta )}{\Gamma \left(\frac{\Delta }{2}\right)^2} 
\int_0^1 \frac{dt}{t(1-t)} X^{\Delta/2}\, _2F_1\left(\frac{\Delta
   }{2},\frac{\Delta }{2};\Delta -\alpha ;X\right)\,.
   \eeq
Now let us use the hypergeometric identity  
\bea
	_2 F_1(a,b;c;x)=(1-x)^{-b}\ _2F_1\left(c-a,b;c;\frac{x}{x-1}\right)\,.
	\label{eq:flip}
	\eea	
The resulting expression factorizes nicely in terms of $z$ and $\bar z$:
\beq
G_{\Delta,0}=\frac{\Gamma (\Delta )}{\Gamma \left(\frac{\Delta }{2}\right)^2} 
\int_0^1 \frac{dt}{t(1-t)} Y^{\Delta/2}
\, _2F_1\left(\frac\Delta2,\frac\Delta2-\alpha ;\Delta -\alpha;-Y\right)\,,
     \eeq
\beq
Y\equiv \frac{X}{1-X}= \frac{ t(1-t) z \bar z}{(1-t z)(1-t \bar z)}\,.
\eeq
Now replace ${}_2F_1$ by its defining power series expansion in $(-Y)$ and integrate the series term by term. For $z=\bar z$, the resulting integrals are of the form \reef{eq:doublesum} and give hypergeometric functions $\, _2F_1(\Delta+2n,\Delta/2+n;\Delta+2n ;z)$, which are 
elementary.\footnote{For $z\ne \bar z$ we would have obtained a series in Appel $F_1$ functions.}
 So we get:
\beq
G_{\Delta,0}|_{z=\bar z} =\left(\frac{z^{2}}{1-z}\right)^{\Delta/2}\sum_{n=0}^\infty
\frac{[(\Delta/2)_n]^3(\Delta/2-\alpha)_n}{n! (\Delta )_{2n} (\Delta -\alpha )_n} \left(\frac{z^{2}}{z-1}\right)^n \,.
   \label{eq:zgen}
   \eeq
Expressing $ (\Delta )_{2n}$ via the duplication formula for the $\Gamma$ function, the series is recognized to be of the ${}_3F_2$ type, and we get precisely Eq.~\reef{eq:3F2-0}.

Is there a similar closed form representation for generic unequal external dimensions or, more specifically, for generic $\Delta_{12}=\Delta_{34}\ne0$ (as would be needed for the crossing symmetry analysis of the $\langle \sigma \en\sigma\en\rangle$ correlator)? The following reasoning shows that this may be difficult. For $D=2$, Eq.~\reef{eq:3F2-0} can be derived starting from the explicit expression \reef{eq:DO}, passing to the variable ${z^2}/({4 (z-1)})$ via the identity
	\beq
	\textstyle{}_2 F_1(a,b,2b,z)=\left(1-\frac z2\right)(1-z)^{-\frac{a+1}2}  \ _2F_1\left(\frac{1-a+2b}2,\frac{a+1}2,b+\frac 12; \frac{z^2}{4(z-1)}\right)\,,
	\label{eq:C4}
	\eeq
and then aiming for Clausen's formula (\cite{Bateman}, Sec.~4.3) to express the square of a ${}_2F_1$ as a ${}_3F_2$. However, Eq.~\reef{eq:C4} is not useful for generic unequal dimensions.

Passing to the spin 1 case, the idea is to use the second recursion relation \reef{eq:rec2} which expresses spin 1 blocks via the spin 0 ones. This relation can be restricted to the $z=\bar z$ line, as the differential operator $\calF_1$ acts within the line; for $l=1$ it gives
\beq
G_{\Delta, 1}(z)=\frac1{2(\Delta-\alpha)}\left[ \frac{2-z}{2z}(\Delta-2\alpha-1)
+(1-z) \del_z\right]G_{\Delta+1,0}(z)\,.
\eeq
 Substituting the spin 0 closed form expression \reef{eq:3F2-0}, we find 
\beq
G_{\Delta, 1}(z)=\frac{2-z}{2(\Delta-\alpha)z}
\left(\frac{z^2}{1-z}\right)^\frac{\Delta+1}2
[ y\,\del_y+\Delta-\alpha]f(y)\,,
\eeq
 where $f(y)$ (with $y\equiv{z^2}/{4 (z-1)}$) is the ${}_3F_2$ function entering the expression for $G_{\Delta+1,0}(z)$. Eq.~\reef{eq:3F2-1} then follows, since the ${}_3F_2$ function satisfies 
 \beq
[ y\del_y+b_2-1] {}_3F_2(a_1,a_2,a_3;b_1,b_2;y)=(b_2-1){}_3F_2(a_1,a_2,a_3;b_1,b_2-1;y)\,.
\eeq

\section{Recursion Relation for the Transverse Derivatives}
\label{app:CKrec}

The following recursion relation for $h_{m,n}$ can be derived by applying $\del_a^m\del_b^n$ to the Casimir equation~\reef{eq:Cas} written in the $a,b$ coordinates, setting $a\to1$, $b\to0$, and shifting $n\to n-1$:
\begin{align}
 &2(D+2n-3)h_{m,n}=\nn\\
 &\hspace{1cm}2m (D+2 n-3)[-h_{m-1,n}+(m-1) h_{m-2,n} +(m-1)(m-2)h_{m-3,n}] \nn\\
 &\hspace{1cm}-\,\,h_{m+2,n-1} + (D-m-4 n+4) h_{m+1,n-1}\nn\\
 &\hspace{2cm}+\,\, \left[2 C_{\Delta ,l}+2 D (m+n-1)+m^2+8m n-9m+4 n^2-6
   n+2\right]h_{m,n-1} \nn\\
   &\hspace{2cm}+\,\,m \left[D (m-2 n+1)+m^2+12 mn-15m+12 n^2-30 n+20\right] h_{m-1,n-1} \nn\\
 &\hspace{1cm}+\,\,(n-1) [h_{m+2,n-2}-(D-3 m-4 n+4) h_{m+1,n-2}]\,. 
  \end{align}

\section{Linear Programming Implementation}
\label{app:details}

\newcommand\De{\Delta}
\newcommand\de{\delta}
\newcommand\s{\sigma}

Let us write the crossing constraint Eq.~(\ref{eq:cross}) as
\beq
\label{eq:crossingconstraintrewritten}
0=F^{\De_s}_{0,0}(u,v)+\sum \nolimits' p_{\De,l}F^{\De_\s}_{\De,l}(u,v)\,,
\eeq
where $F^{\De_\s}_{\De,l}(u,v)\equiv v^{\De_\s}G_{\De,l}(u,v)-u^{\De_\s}G_{\De,l}(v,u)$.  To rule out some spectrum of operator dimensions, it suffices to find a linear functional $\Lambda$ acting on functions of $(u,v)$ such that
\begin{enumerate}
\item $\Lambda(F^{\De_\sigma}_{0,0})=1$ (normalization condition)
\item $\Lambda(F^{\De_\sigma}_{\De,l})\geq 0$ for all $\De,l$ in the spectrum (positivity constraints).
\end{enumerate}
Any such $\Lambda$ would be inconsistent with the crossing relation Eq.~(\ref{eq:crossingconstraintrewritten}) and positivity of the coefficients $p_{\De,l}$, implying that the putative spectrum cannot be realized in a unitary (or reflection positive) CFT.

In practice, we consider $\Lambda$ of the form
\bea
\Lambda:F(u,v) 
&\mapsto& \sum_{m+2n\leq 2n_\mathrm{max}+1}\lambda_{m,n}\partial_a^m\partial_b^nF(a,b)|_{a=1,b=0} 
\eea
where the variables $a,b$ are defined in Eq.~(\ref{eq:abdefinition}), $\lambda_{m,n}$ are real coefficients, and the range of $m,n$ depends on an integer $n_\mathrm{max}$.  Since $F^{\De_\s}_{\De,l}(u,v)$ is antisymmetric under $u\leftrightarrow v$, only odd $a$-derivatives are nonzero, and a given $n_\mathrm{max}$ corresponds to $(n_\mathrm{max}+1)(n_\mathrm{max}+2)/2$ nonzero coefficients $\lambda_{m,n}$.  Larger $n_\mathrm{max}$ gives stronger bounds, but is more computationally intensive.  Derivatives of $F^{\De_\s}_{\De,l}$ are simply linear combinations of derivatives of the conformal blocks $G_{\De,l}$, which we compute in \texttt{Mathematica} using the methods outlined in Section~\ref{sec:blocks}.
We first evaluate the derivatives $\partial_a^mG_{\De,l}$ up to $m=2n_\mathrm{max}+1$ and all the other derivatives in the range $m+2n\leq 2n_\mathrm{max}+1$ follow via the recursion relation of Appendix \ref{app:CKrec}. To compare with previous work, we have $n_\mathrm{max}=N/2=k-1$ where $N$ and $k$ are the parameters used in \cite{Rychkov:2009ij} and \cite{Poland:2011ey}, respectively.

We implement the positivity constraints above by discretizing the set of dimensions and restricting the spin to lie below some large finite value.  The conformal blocks $G_{\De,l}(u,v)$ converge quickly for large dimensions and spins, so this is a reasonable approximation.  It can be made arbitrarily good by using finer discretizations and a larger maximum dimension and spin.  The plots in this paper were generated with the choices given in Table~\ref{tab:cbderivtables}.
\begin{table}[h]
\centering
\begin{tabular}{|c|c|c|c|}
\hline
 & $\de$ & $\De_\mathrm{max}$ & $L_\mathrm{max}$\\
\hline\hline
T1 & $2\times 10^{-5}$ & 3 & 0\\
\hline
T2 & $5\times 10^{-4}$ & 8 & 6\\
\hline
T3 & $2\times 10^{-3}$ & 22 & 20\\
\hline
T4 & 0.02 & 100 & 50\\
\hline
T5 & 1 & 500 & 100\\
\hline
\end{tabular}

\caption{In this work, we used a combination of five tables T1-T5 of conformal blocks (and their derivatives) with different discretizations, maximum dimensions, and maximum spins.  For each table, dimensions were chosen from the unitarity bound $\De_\mathrm{min} \equiv l+1-\frac12 \delta_{l,0}$ to $\De_\mathrm{max} + 2 (L_{\mathrm{max}} - l)$ with step $\delta$, and spins were restricted to $0\le\l\leq L_\mathrm{max}$.  The choices above allow for high-resolution studies of the low-spin spectrum (T1-T3), while simultaneously ensuring control of intermediate dimensions and spins (T4), and also asymptotic behavior (T5).\label{tab:cbderivtables}}
\end{table}

After restricting the dimensions and spins to lie in a finite set, our problem becomes a standard linear programming problem which can be solved on a computer.  Solvers are available in a wide variety of software libraries and applications, including for example \texttt{Mathematica}.   Here, we choose to use the dual simplex algorithm implementation in IBM's \texttt{ILOG CPLEX Optimizer}.\footnote{
\href{http://www-01.ibm.com/software/integration/optimization/cplex-optimizer/}{http://www-01.ibm.com/software/integration/optimization/cplex-optimizer/}}

To generate plots like those in Figure~\ref{fig:deltaE-Eprime-above}, we must scan over different choices of dimensions $\De_\sigma,\De_\epsilon,\dots$, solving a linear program each time to determine the boundary between feasible and infeasible choices.  When scanning over a single dimension, for example, this is most efficiently done using a binary search.  One can additionally generalize binary searches to work in higher dimensions by recursively refining a lattice of points.  These algorithms are readily parallelizable, and it is very convenient to take advantage of a cluster of machines to perform the computations.  Our search logic is implemented in \texttt{Scala}, taking advantage of its actor model for distributing parallel tasks across a network, and \texttt{ILOG CPLEX}'s \texttt{Java} (\texttt{Scala} compatible) API for performing the computations.

\bibliography{Biblio}{}
\bibliographystyle{utphys}

\end{document}